\documentclass[adp,fleqn]{w-art}
\usepackage{times}
\usepackage{w-thm}
\usepackage{citesort}
\usepackage[]{graphicx}
\chardef\bslash=`\\ 

\newcommand{\ve}[1]{\text{\boldmath $#1$}}
\newcommand{\avg}[1]{\langle #1 \rangle}
\newcommand{\dd}{\text{d}}

\begin{document}

\DOIsuffix{theDOIsuffix}
\Volume{14}
\Issue{1}
\Copyrightissue{01}
\Month{01}
\Year{2005}
\pagespan{1}{}


\title[Brownian motion in soft matter and biological physics]{Brownian
  motion: \\ a paradigm of soft matter and biological physics}

\author[E. Frey]{Erwin Frey\inst{1,3},\footnote{\; Corresponding author\;
        E-mail: frey@lmu.de}} 
\author[K. Kroy]{Klaus Kroy\inst{2,3},\footnote{E-mail: kroy@hmi.de}}

\address[\inst{1}]{Arnold Sommerfeld Center and CeNS, Department of
  Physics, Ludwig--Maximilians--Universit\"at M\"unchen,
  Theresienstrasse 37, 80333 M\"unchen, Germany}

\address[\inst{2}] {Institut f\"ur Theoretische Physik, Universit\"at
  Leipzig, Linnestr. 5, 04103 Leipzig, Germany}

\address[\inst{3}] {Hahn--Meitner
  Institut, Glienicker Strasse 100, 14109 Berlin, Germany}


\begin{abstract}
  This is a pedagogical introduction to Brownian motion on the
  occasion of the 100th anniversary of Einstein's 1905 paper on the
  subject. After briefly reviewing Einstein's work in its contemporary
  context, we pursue some lines of further developments and
  applications in soft condensed matter and biology.  Over the last
  century Brownian motion became promoted from an odd curiosity of
  marginal scientific interest to a guiding theme pervading all of the
  modern (live) sciences.
\end{abstract}

\maketitle                   

\section{Introduction: 100 years ago}
``If, in some cataclysm, all of scientific knowledge were to be
destroyed, and only one sentence passed on to the next generation of
creatures, what statement would contain the most information in the
fewest words? I believe it is [\dots] that \emph{all things are made
of atoms --- little particles that move around in perpetual motion,
attracting each other when they are a little distance apart, but
repelling upon being squeezed into one another.}'' Some decisive
theoretical tools to put the \emph{atomic hypothesis} (here quoted
from \emph{The Feynman Lectures} \cite{feynman_lectures:I}) at
scrutiny were provided in Einstein's paper ``\"Uber die von der
molekulartheoretischen {Theo\-rie} der W\"arme geforderte Bewegung von
in ruhenden Fl\"ussigkeiten suspendierten Teilchen''
\cite{einstein:05} published in this journal 100 years ago, which
starts with the claim ``that according to the molecular-kinetic theory
of heat, bodies of a microscopically visible size suspended in liquids
must, as a result of thermal molecular motions, perform motions of
such magnitudes that they can be easily observed with a microscope.''
It quickly became clear that Einstein's calculations could explain
hitherto mysterious observations of dancing pollen granules and other
small particles first systematically described by the Scottish
botanist Robert Brown \cite{brown:28} and (independently) by his
French colleague Adolphe Brongniart \cite{brongniart:27}. And a few
years later, drawing upon painstaking observations of thousands of
Brownian particles in his lab at the Sorbonne, Jean Perrin could
ultimately end the debate about the reality of atoms
\cite{perrin:nobel}. Though dating back to the ancient Greeks, this
debate had considerably intensified towards the turn of the century
and prompted Planck's dictum that the acceptance of new scientific
ideas proceeds via natural extinction rather than by persuasion of the
opponents, which is sometimes referred to as \emph{Planck's principle}
among philosophers of science.

It was soon realized that the laws of Brownian motion do not
exclusively pertain to the realm of physics but likewise to chemistry,
biology and even economics.  After the dust had settled, this was
acknowledged by a couple of Nobel prizes: in physics, Perrin was
rewarded for proving the reality of atoms while the chemists at the
same time honored Svedberg for his dispersion studies and for the
invention of the ultracentrifuge, and later Onsager for further
growing the germ layed in Einstein's 1905 paper. Also, the economists
eventually embraced the laws of Brownian motion as useful tool for the
valuation of stock options in due time for the centenary of their
derivation by Bachelier \cite{bachelier:00}. Since he (and other
contributors) had unfortunately passed away by that time, the prize
went to Merton and Scholes\footnote{Commonly better known for their
role in one of the world's largest financial collapses
\cite{dunbar:2000}.}.

{F}rom the historical perspective, the theory brought forward in
Ref.~\cite{einstein:05} (together with related work by Smoluchowski
\cite{smoluchowski:06} and Langevin \cite{langevin:08}) is therefore
certainly not to be sniffed at by her two sisters, \emph{relativity}
and \emph{quantum mechanics}. It took quite a different course,
though. While the latter two were soon promoted to the holy grail of
science and have never since ceased to attract the public, the former
catched much less attention. This is all the more astonishing as the
vision it promoted, namely that our macroscopic world including its
living organisms resides on a jittering sea of fluctuations, is
particularly remarkable for the broadness of its impact. During the
last 100 years, the facets of this vision have continuously
multiplied, thereby providing ample evidence for a ``slower''
\cite{haw:2005}, yet ``unfinished'' \cite{coleman:2003} revolution.
Brownian motion remains a leading paradigm in a development that seems
directed oppositely to that of the quantum $\&$ relativity revolution
--- not deeper into the elements but towards the understanding of the
principles underlying the emergence of collective properties in
many--particle assemblies\footnote{In view of beautiful recent
  progress in what one might call ``condensed matter cosmology''
  (e.~g.\ \cite{antunes-bettencourt-zurek:99,volovik:2001}), this
  antithesis seems less compelling now than some years ago.} --- and
for which P. W. Anderson has coined the mantra ``more is different''
\cite{anderson:72}, meaning that ``at each new level of complexity,
entirely new properties appear, and the understanding of these
behaviors requires research which I think is as fundamental in its
nature as any other.''  Depending on the importance of thermal
fluctuations one may roughly divide this research into ``hard matter''
(``$k_BT\to0$'') and ``soft matter''
\cite{chaikin-lubensky:95,cates-evans:2000} (``$k_BT\simeq
1$'')\footnote{Here, the characteristic intrinsic energy scale of the
  system is understood to provide the unit. In this sense,
  $k_BT\approx 1$ may for instance indicate that Brownian motion
  limits our auditory abilities \cite{denk-watt:89}, or the accuracy
  of a gravitational wave detector \cite{saulson:90} or (in the form
  of Nyquist--Johnson noise) of any electrical measuring instrument.}
physics.  Underlying both is the discipline of statistical mechanics,
which has not only become a major part of any physics curriculum, but
which has over the years proven its vitality by invading numerous
foreign disciplines such as biology, ecology, economics, traffic,
sociology, game theory, etc.
\cite{kerner:2004,schweitzer:2003,challet-etal:2001,%
mantegna-stanley:2000,bouchaud-potters:2004,%
neumann_morgenstern:book}, and by identifying new interdisciplinary
research topics like criticality, structure formation and synergetics,
quenched disorder, glassy dynamics, and many others, where it tries to
bring to bear its concepts and methods.  On the other hand, widespread
hopes that on a more fundamental, sub--atomic scale fluctuations would
eventually cease and give way to solidity and firm order did not quite
materialize. Instead, the toolbox of statistical mechanics turned out
to be instrumental also to various ``fundamental'' developments from
quantum theory to cosmology (path integrals, stochastic quantization,
space--time foam, \dots).

In the following, our scope will have to be much, much narrower,
though, and we will try to restrict ourselves to tracing a selective
set of implications and modern offsprings of Einstein's 1905 paper on
Brownian motion \cite{einstein:05}, with the aim of illustrating
variations of its basic themes in soft condensed matter and biological
physics. Since, above all, Brownian motion is particularly strongly
linked to biology, not only by its name; a link that we welcome as a
guide (or excuse) for taking shortcuts through the bewildering jungle
of applications.  Soft matter and biological physics are both fields
governed by meso--scale structures for which thermal fluctuations are
all important, the relevant dynamics often bridging the gap
between the microscopically fast and the macroscopically slow, as
neatly summarized by the formulas
\begin{equation}\label{eq:kT}
 k_BT\approx 4\text{ pN$\cdot$nm}\;, \qquad
 k_BT/\eta_{\text w} \approx 4 \text{
 $\mu$m$^3$s$^{-1}$}\;,
\end{equation}
where $\eta_{\text w}$ denotes the viscosity of water. Conversely,
these formulas immediately raise the (legitimate) expectation that
Brownian motion is an ubiquitous phenomenon in these fields.

More precisely, our outline is as follows. After a brief historical
introduction to Einstein's paper \cite{einstein:05} and closely
related work in section~\ref{sec:history}, the remainder is organized
along six traces. Our starting point is the observation that the
derivation in Ref.~\cite{einstein:05} inspired and partly anticipated
major developments in non-equilibium statistical mechanics
(section~\ref{sec:gse}) as well as a powerful stochastic calculus
(section~\ref{sec:methods}) with innumerable applications, of which we
only mention a few paradigmatic examples taken from biology in
section~\ref{sec:noisy_world}.  Precisely as proposed by Einstein in
1905 \cite{einstein:05}, Brownian particles have ever since served as
versatile probes of their environment (section~\ref{sec:probe}).
Moreover, their stochastic fluctuations have provided valuable
insights into their own (soft) internal degrees of freedom, which can
give rise to most complex macroscopic material properties thus
providing a unique potential for applications
(section~\ref{sec:fluctuating_manifolds}). Finally, Brownian
fluctuations are both exploited and rectified in ingenious biological
micro--machines (section~\ref{sec:motors}).

Needless to say that, even within this limited scope, we cannot be
objective or complete in any sense. Original publications can by no
means be cited systematically, and citations should rather be taken as
our (certainly biased) personal suggestions for further reading. With
these cautious remarks, we hope that the reader is prepared for a very
subjective random walk lurching to and fro between physics and
biology.

\section{Historical notes: from Brown to Perrin}\label{sec:history}
\subsection*{\bf Brownian motion: a very brief history}
``The story of Brownian motion is one of confused experiment, heated
philosophy, belated theory, and, finally, precise and decisive
measurement'' \cite{haw:2002}. It is well documented in various
publications by Perrin, Nye, Brush, and others, and summarized in
several monographs on Brownian motion (e.~g.\
\cite{chandrasekhar:43,nelson:67,gardiner:85,mazo:2002}). Here, we
content ourselves with a few remarks. First, since we expect 2005 to
be a year with strong ``Matthew effects'' (in case of doubt, credit is
ascribed to those who already got most \cite{mermin:2004}), it seems
in order to start by a brief disclaimer: Brown was not the first
observer of Brownian motion (similar observations had repeatedly been
made after the first microscopes came up in the late 17th century);
Perrin was not the first to show that kinetic theory is in accord with
experiments, nor was the admirable diligence of his group necessary to
make the atomic hypothesis acceptable to the majority of researchers;
Einstein did of course not invent the diffusion equation --- which he
supposed to be generally known just as we would today --- nor did he
first suggest that Brownian motion is due to thermal motion of the
solvent molecules or provide the first testable prediction of
statistical mechanics or kinetic theory, etc.

In fact, Brown was curious enough about the erratic dance under the
microscope to spend some time in the late 1820's on systmatic
investigations, trying to clarify its origin. Starting out with
micrometer--sized granules from the pollen of \emph{Clarkia pulchella}
and later using several other substances, he convinced himself (though
not everybody else) that internal animate as well as external
perturbative causes and optical illusions could be ruled out.
Nevertheless, the phenomenon hardly received the attention it
deserved.  It remained a curiosity during Brown's lifetime and almost
until the end of the century, when L\'eon Gouy's detailed
investigations of the phenomenon \cite{gouy:88} led him to the
conclusion that the ubiquitous perpetual motion is caused by thermal
motion of the solvent molecules. (This was in 1888, when Einstein
still went to primary school, and even at that time the idea was not
really new\footnote{See e.~g.\ C. Wiener, ``Erkl\"arung des
atomistischen Wesens des tropfbar--fl\"ussigen K\"orperzustandes und
Best\"atigung desselben durch die sogenannten Molecularbewegungen''
\cite{wiener:63}.}) He later also noticed that this explanation seemed
to be in contradiction to the second law of thermodynamics (a concern
independently raised by R{\"o}ntgen in a letter to Einstein in 1906)
or could at least provide an ideal \emph{natural laboratory} for
direct experimental examination of certain apparent logical
contradictions between the atomic hypothesis and thermodynamics.  By
that time diffusion was a well--established thermodynamic notion
routed in the work ``\"Uber Diffusion'' (1855) by the German
pathologist Adolf Fick, which starts with the words
``Hydrodiffusion through membranes is not only one of the basic
factors of organic life\footnote{Some advocates of the opposite view
nowadays claim to have compelling evidence that ``there is certainly
no reason to suppose that classical diffusion theory, or any of its
offspring, will play a significant role in our understanding of
biology in the future''. \cite{agutter-malone-wheatley:2000}}, but
also a most interesting physical process that deserves much more
attention from physicists than it has got so far''
\cite{fick:1855}. But the relation between Brownian motion and
diffusion --- today often regarded as synonymous --- was much less
clear. This was the missing link provided by Einstein and
Smoluchowski, who thereby pointed the way to \emph{quantitative}
hands--on experiences with a strange microcosmos awaiting
discovery. Despite of work by Clausius, Maxwell, Boltzmann, Loschmidt
and others on heat conduction and interdiffusion that had clearly
demonstrated the usefulness of kinetic theory and even produced
reasonably accurate estimates for the size of molecules by the late
19th century, there remained strong opposition against taking the
underlying atomistic view too literal. Only Perrin's extensive
studies, performed with Einstein's prediction at hand, could finally
convince the last of the ``Energetiker'' around Ostwald that atoms
were more than just a convenient analogy. To us, the great impact of
Perrin's and Einstein's work on Brownian motion seems partly linked to
the fact that, instead of only measuring another thermodynamic or
hydrodynamic relation that could be rationalized by kinetic theory, a
forthright manifestation of molecular chaos, namely the predicted
``strange'' equation of motion of individual diffusing colloids (on
average the travelled distance grows with the square root of time),
\emph{could directly be observed}. The fractal trajectory implied by
Einstein's result provided a convincing explanation why earlier
attempts to measure the \emph{velocity} of Brownian particles and
compare it to the average thermal velocity predicted by kinetic theory
(for free particles), were in vain: with the characteristic distance
travelled by the particle growing as $\sqrt{t}$, the velocity is
proportional to $1/\sqrt{t}$.  Naturally, around 1900, a
non--differentiable trajectory was hard to swallow, and Einstein
hastened to add \cite{einstein:06} that the molecular discreteness
provides a lower cutoff to such unseemliness.  Curiously, the same
fractal trajectories were determined to reappear later in the disguise
of the Heisenberg uncertainty relation, which indicated that they also
prevail on a more fundamental level \cite{feynman-hibbs:65} (see
section~\ref{sec:methods}).

Finally, it should be said that Einstein's interest in Brownian motion
grew out of a profound interest in statistical fluctuations
\cite{einstein:06,einstein:07,einstein:10}, which can be identified as
a common denominator of his work in seemingly unrelated areas.  In
particular, fluctuations played a crucial role \cite{irons:2004} in
his persistent efforts to find a rational explanation for Planck's
radiation law.  Einstein realized that stochastic fluctuations rule on
the \emph{quantum level} (photon statistics) in classical \emph{small
systems} (Brownian motion) and near a \emph{critical point} (critical
opalescence), thereby anticipating at the beginning of the 20th
century, what became vast fields of research towards its end. Without
exaggerating, he may therefore be called the father of fluctuation
theory \cite{fluctuation-phenomena:87}.

\subsection*{\bf Revision course: Einstein's 1905 paper on Brownian motion}
In brief, the argument in Ref.~\cite{einstein:05} goes as follows. If
the atomic hypothesis holds, particles immersed in an ambient medium
must undergo perpetual irregular motion due to chaotic collisions with
its molecules. It was Einstein's insight (cited above) that
micrometer--sized particles are both big enough to be seen themselves
in the microscope and small enough that their Brownian motion is
substantial, and, as a consequence, keeps them suspended indefinitely
against gravity.  The publication consists of two major parts. In the
\emph{first part}, Einstein argues against the view --- still common
at that time --- that suspended particles may be dismissed in
thermodynamics. He insists that $N$ particles of arbitrary volume $v$
suspended in a solvent at small number density $n\ll v^{-1}$ represent
an ideal gas and therefore give rise to an osmotic pressure $p=nk_BT$
(in modern notation), thereby excerting a finite, albeit small,
thermodynamic force on their container. In presence of a constant
volume force $\ve K$ acting on the particles one thus has the
(isothermal) force balance $n(\ve r)\ve K+\ve\nabla p(\ve r)=0$, or
$p(\ve r)= k_BT n(\ve r)\propto e^{- \ve K \cdot \ve r/k_BT}$ in the
steady state. Further, Einstein suggests to regard this stationary
balance as a dynamic equilibrium between a diffusion current $-D\ve
\nabla n$ and a drift current $n\ve K/\zeta$, with $\zeta$ and $D$ the
friction and diffusion coefficient of the Brownian particles in the
solvent, respectively. Eliminating $\ve K$ and $n$ between the balance
of forces and that of drift and diffusion, he infers what is today
known as \emph{Stokes--Einstein relation}
\begin{equation}\label{eq:ser}
D= k_BT/\zeta \;. 
\end{equation}
So far, this is a relation between two coefficients appearing in two
generalized hydrodynamic equations, namely the Stokes equation for
incompressible viscous flow\footnote{This is nothing but the diffusion
equation, Eq.~(\ref{eq:deq}), with $D$ replaced by the kinematic
viscosity, and $n(\ve r)$ replaced by the solvent velocity field
projected onto its ``transverse'' components (so that it is
non--divergent).}, and the diffusion equation. This is not yet quite,
what made it so remarkable. We must point out that the notion of
Boltzmann's constant $k_B$ was not yet fully established at the time,
and its apparence in Eq.~(\ref{eq:ser}), originally in the form of the
gas constant divided by Avogadro's number, allowed the latter to be
determined from measuring macroscopic quantities. Intriguingly, it
thereby indicated a relation of the macroscopic kinetic coefficients
$D$, $\zeta$ to the microscopic, molecular world.

This link, which was later formalized under the name of
\emph{Green--Kubo relations}, is established in the \emph{second part}
of Einstein's paper, which is quoted as a pedagogical introduction to
stochastic processes in Gardiner's textbook \cite{gardiner:85}. It
undertakes a probabilistic derivation of the diffusion equation based
on a particularly simple model for the fluctuations of the Brownian
particle, called a random walk. The essential idea is to express the
particle concentration (in one space dimension) at position $x$ and
time $t+\tau$ as a function of the concentration at time $t$,
\begin{equation}\label{eq:cke}
n(x,t+\tau)=\int \dd\xi \; n(x+\xi,t)\varphi_\tau(\xi) \;.
\end{equation}
Due to the use of the auxiliary ``jump propabilities''
$\varphi_\tau(\xi)$ in this \emph{Chapman--Kolmogorov equation}, as it
is called today, Einstein's derivation appears much more elegant and
transparent than Smoluchowski's more mechanistic discussion of
Brownian motion, because it circumvents the construction of an
explicit \emph{microscopic dynamical theory} of the process by some
(implicit) technical assumptions on $\varphi_\tau(\xi)$.  Taylor
expanding $n(x,t)$ with respect to the ``small'' quantities $\tau$ and
$\xi$ and matching the leading order terms on both sides, Einstein
recovers the diffusion equation
\begin{equation}\label{eq:deq}
\partial_t n(\ve r,t)=D\nabla^2 n(\ve r,t)
\end{equation}
along with a stochastic expression for
the diffusion coefficient,
\begin{equation}\label{eq:msd}
 2 D=\frac1\tau \int \dd\xi\; \xi^2 \varphi_\tau(\xi) =
\frac{\avg{\delta x^2}}\tau =\frac{\avg{\delta \ve r^2}}{3 \tau }\;.
\end{equation}
The expressions for $D$ are understood to be independent of the
considered time interval $\tau$ as long as $\tau$ is large compared to
a microscopic collision time. In his derivation Einstein also invokes
the limit ``$\tau\to 0$'', which later became the subject of debate
(see below).  The final relation between $D$ and the mean square
displacement $\avg{\delta\ve r^2(t)}\equiv \avg{[\ve r(t) -\ve
r(0)]^2}$ of a Brownian particle starting at $t=0$ at $\ve r=\ve r(0)$
follows from applying the diffusion equation to each of the diluted
particles separately. It is of interest as paradigm for a general
rule: kinetic coefficients (here $D$, $\zeta$) in macroscopic
thermodynamic or hydrodynamic equations (here diffusion, Stokes
equation) can entirely be expressed in terms of correlation functions
of fluctuations of microscopic variables (see
section~\ref{sec:gse}). This is how the discrete chaotic processes on
the microscale generate the smooth behavior on the macroscopic scale.

\section{Non-equilibrium statistical mechanics: from Stokes--Einstein
  relations to fluctuation theorems and effective temperatures}
\label{sec:gse}

\subsection*{\bf Close to equilibrium: the fluctuation--dissipation
  connection} 

The most celebrated result of Eintein's 1905 paper on Brownian motion
\cite{einstein:05} is the so--called Stokes--Einstein relation that
links two hydrodynamic transport coefficients to each other and to
thermal fluctuations, see Eqs.~(\ref{eq:ser}),~(\ref{eq:msd}). The
second part of it, the relation between hydrodynamic transport
coefficients and microscopic fluctuations, is today more commonly
known as \emph{Green--Kubo relation}, and is usually presented in the
form \cite{kubo:86}
\begin{equation}\label{eq:gkr}
  D = \frac13\int_0^\infty \!\!\dd t \;\avg{\ve v(t)\cdot \ve v(0)} \;.
\end{equation}
As Einstein himself noticed \cite{einstein:06,einstein:07}, the
content of Eq.~(\ref{eq:gkr}) is not restricted to Brownian motion but
is straightforwardly translated to electrical circuits\footnote{Since
he could not imagine other testable consequences, he dismissed any
discussion of further realizations as ``useless'' \cite{einstein:06}
but maintained interest in the electrical analog \cite{einstein:07},
for which he even designed a ``Maschinchen'' (little machine)
\cite{foelsing:97} to do experiments by himself.}, a case that was
later further elaborated by Johnson and Nyquist
\cite{johnson:28,nyquist:28}. Analogous relations have been derived
for all kinds of transport coefficients (conductivities, viscosities,
etc.) relevant in different areas of condensed matter physics. They
are of considerable practical interest, as they suggest a way to do
non--invasive measurements of response coefficients (here
$\zeta$). And they provide insight into the complicated behavior of
strongly interacting many--particle systems. For example, the general
form of Eq.~(\ref{eq:gkr}) persists for dense suspensions of Brownian
particles, where the gradient diffusion coefficient governing the
diffusive spread of concenrtation variations via the diffusion
equation becomes renormalized by the (osmotic) compressibility of the
suspension compared to the bare diffusion coefficient of a single
particle in the solvent. This knowledge allows one (among other
things) to infer why the dynamics slows down near phase transitions,
where susceptibilities (here the compressibility) diverge.  Thus, near
a critical point \cite{domb:96}, fluctuations not only become
long--ranged as foreseen by Einstein in his paper on critical
opalescence \cite{einstein:10}, but they also become long--lived;
though in a slightly more intricate way than suggested here
\cite{hohenberg-halperin:77,frey_schwabl:94}.

It was Onsager who started the excavation of the treasure buried in
the Stokes--Einstein relation. Following Einstein's later work
\cite{einstein:10} on fluctuations, he likened the relaxation of
hydrodynamic variables in response to weak external perturbations to
the decay of correlations between the corresponding microscopic
variables: the forced deviation from equilibrium could as well have
been a random fluctuation of the system.  This \emph{regression
principle} allowed him to exploit exact microscopic symmetries,
notably the invariance of the microscopic equations of motion under
time reversal, to infer constraints on the kinetic coefficients:
however difficult and incomprehensible e.~g.\ the coefficient of
thermodiffusion (Soret effect)\footnote{See e.~g.\
Ref.~\cite{wiegand:2004} for a recent experimentally oriented review,
and various contributions in Eur.~Phys.~J.~E \textbf{15}(1), 2004.} may seem,
thanks to Onsager we can exactly relate it to a corresponding
coefficient for ``diffusive heating'' (Dufour effect). Further
generalizations of Eq.~(\ref{eq:ser}) like the mentioned Green--Kubo
relations and the fluctuation--dissipation theorem followed.  Around
these cornerstones the field of linear non-equilibrium thermodynamics
(also ``irreversible thermodynamics'', ``generalized/molecular
hydrodynamics'') developed. The modern derivation
\cite{kadanoff-martin:63,forster:75,brenig:89} of all of the mentioned
relations is based on \emph{linear--response theory} (often
complemented by projection operator techniques \cite{zwanzig:2001} to
separate slow, hydrodynamic from fast, fluctuating variables), which
therefore can be regarded as the foundation of
``close--to--equilibrium statistical mechanics''; see Kubo's short
pedagogical overview \cite{kubo:86}.  The basic assumption underlying
linear--response theory is that the phase--space density may be
linearized in a small perturbation added to the Hamiltonian of the
system. This has been the subject of some debate, reviewed in
\cite{dorfman:99}: since single trajectories of nonlinear dynamical
systems are known to exhibit sensitive dependence on perturbations,
the required macroscopic linearity cannot be a property of single
trajectories but rather has to rely on some ergodic property of the
phase space dynamics as a whole
\cite{gallavotti-bonetto-gentile:2004}. Only for toy models such as
the periodic Lorentz gas or Sinai billiard \cite{bunimovich-sinai:81},
the link between the transport coefficients and the underlying
microscopic Hamiltonian dynamics can be made fully explicit
\cite{gaspard:98,dorfman:99}, while for the complex systems generally
of interest in statistical mechanics this is out of reach. Already a
rigorous microscopic derivation of Fourier's law for heat conduction,
which is a diffusion equation with $D$ replaced by the thermal
conductivity, turns out to be quite an endeavor (most recently
reviewed in Ref.~\cite{bonetto-lebowitz-rey_bellet:tbp}). Similarly,
it was recently disputed (see \cite{cecconi-etal:2005}) whether
extremely careful measurements \cite{gaspard-etal:98} might be able to
establish the role played by chaotic microscopic dynamics for Brownian
motion. So the question, whether chaotic microscopic dynamics or the
high dimensionality of the phase space is responsible for the
irregularity of Brownian motion is, after a century, still open for
investigation.
\begin{figure}[htb]
\begin{minipage}[t]{.45\textwidth}
  \includegraphics[width=\textwidth]{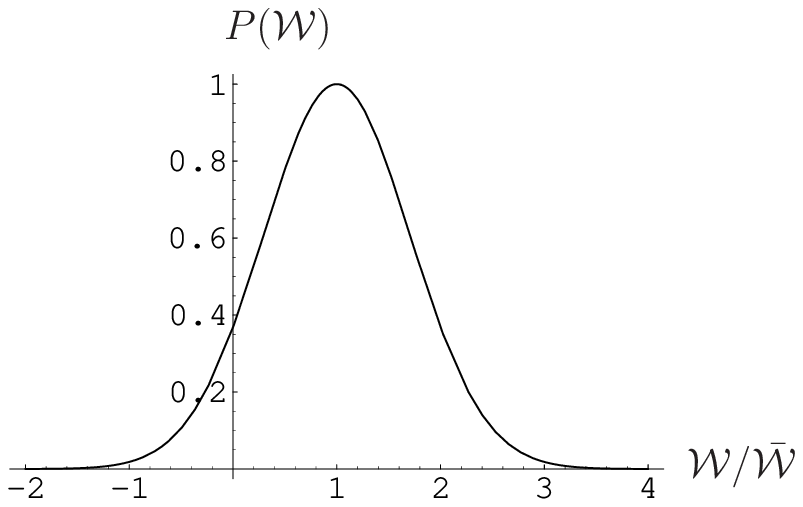}
\caption{Probability distribution for the dissipative work ${\cal W}$.
The finite weight for negative ${\cal W}$ implies that Einstein's
theory of Gaussian fluctuations, if applied to ${\cal W}$, predicts
transient violations of the second law.}
\label{fig:fluctuation_theorem}
\end{minipage}
\hfil
\begin{minipage}[t]{.45\textwidth}
\includegraphics[width=\textwidth]{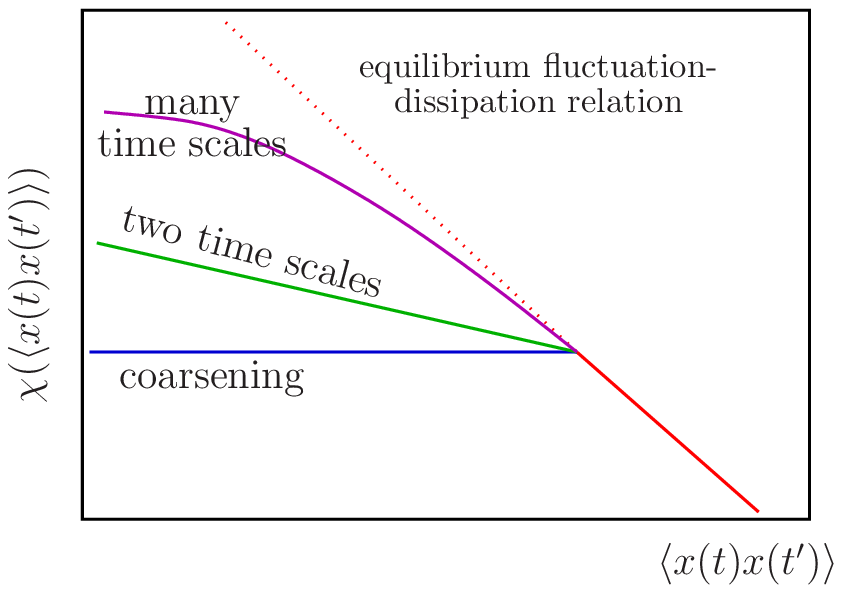}
\caption{Various scenarios for generalized fluctuation-dissipation relations
  represented in integrated form: $\chi=\int_{t'}^{t} \! d\tau
  \partial\avg{x(t)}/\partial F(t')$ }
\label{fig:qfdt}
\end{minipage}
\end{figure}

\subsection*{\bf Far from equilibrium: ``violating'' basic laws and aging} 

Far from equilibrium, the whole formalism of irreversible
thermodynamics is \emph{a priori} not generally applicable. However,
it is a natural temptation to try to extend the approved formalism to
classes of problems that are in some generic sense far away from
equilibrium, such as driven diffusive
systems~\cite{schmittmann_zia:95}, sheared fluids
\cite{berthier-barrat:2002} or gelled \cite{cipelletti-etal:2000}
suspensions of Brownian particles, a living cell
\cite{fabry-etal:2001} or RNA molecules or protein domains unfolding
upon application of an external pulling force
\cite{liphardt-etal:2002}, etc. Can, for such systems, still general
statements be made that save us from having to solve the full
dynamics? One approach tries to extend the classification scheme of
universality classes \cite{hohenberg-halperin:77} based on the
renormalization group and equilibrium critical points to systems far
from equilibrium \cite{taeuber:notes} (see section~\ref{sec:methods}).
Closer related to the topics so far dealt with in the present section
are two interesting lines of investigation that try to generalize the
fluctuation dissipation theorem, which we recall is a generalization
of the Stokes--Einstein and Green--Kubo relations. Both are to date
still controversial and the subject of heated debates
(e.~g.~\cite{jarzynski:2004}).

The \emph{first} is concerned with the intensely discussed paradigm of
\emph{glassy dynamics} \cite{les_houches:2003}. Glassy dynamics is
characteristic not only of structural glasses (such as colloidal,
polymeric, molecular, or metallic glasses) and spin glasses but of a
much wider class including many soft \cite{sollich-etal:97},
disordered \cite{young:97}, granular \cite{buchanan:2003}, and in
particular biological (e.~g.\ proteins \cite{iben-etal:89}) systems.
As generic explanation for the origin of glassy dynamics several
tentative mechanisms have been proposed (purely kinetic models, drop
models, trap models, free--energy landscapes, etc.\
\cite{bouchaud-georges:90,sollich-etal:97,bouchaud-etal:97,debenedetti-stillinger:2001}),
and their further exploration to unveil the intricate roles played by
disorder and fluctuations remains a formidable challenge in
theoretical physics.  In contrast to the dynamics in supercooled
liquids that approach the glass transition from the liquid side, which
is rather well understood (see our discussion towards the end of
section~\ref{sec:probe}), the emphasis here is on the non--equilibrium
dynamics after falling out of equilibrium, which is characterized by
\emph{aging} (i.~e.\ relaxing indefinitely, in general in a
non--universal manner, without ever reaching equilibrium)
\cite{bouchaud-etal:97}. Yet, a generalized fluctuation--dissipation
theorem ($t\geq t'$)
\begin{equation}\label{eq:gfdt}
\frac{\delta \avg{x (t)}}{\delta F(t')} =
  \frac1{k_B{\cal T}(t')}\frac{\partial}{\partial t'}\avg{x(t)x(t')}
\end{equation}
may still hold \cite{bouchaud-etal:97} (see
figure~\ref{fig:qfdt}). Close to equilibrium, ${\cal T}$ is the
ordinary ambient temperature $T$ and both sides depend on the time
difference $t-t'$ only. Equation~(\ref{eq:gfdt}) is then just a
time--dependent generalization of Eqs.~(\ref{eq:ser}),~(\ref{eq:msd})
and the Stokes law $F=\zeta v$: the ordinary equilibrium
fluctuation--dissipation theorem. Far from equilibrium (e.~g.\ below
the glass transition temperature), this is still the case for $t-t'\ll
t'$, where $t'$ is identified with the waiting time after falling out
of equilibrium. For $t-t'\gg t'$ time--translation invariance and the
relation between ${\cal T}$ and the ambient temperature are lost. The
response and fluctuation functions in Eq.~(\ref{eq:gfdt}) are then
dominated by their aging parts. The hope is that \emph{the effective
temperature} ${\cal T}(t')$ will still exist and be independent of the
variable (here $x$) and consistent with alternative definitions of
effective temperatures that are \emph{a priori} independent of
Eq.~(\ref{eq:gfdt})\footnote{The effective temperature may, however,
exhibit a more complicated time dependence of the form ${\cal
T}(\avg{x(t)x(t')})$, as observed in certain spin glasses, or simply
diverge (corresponding to a universal ``trivial'' aging dynamics), as
e.~g.\ in ordinary coarsening during an equilibrium phase
transition.}. This, in fact, would justify the notion of a generalized
or quasi fluctuation--dissipation theorem
\cite{crisanti-ritort:2003}. While the theoretical understanding of
these issues emerged from mean--field spin glass models
\cite{cugliandolo-kurchan:93}, a corresponding development for
structural glasses (e.~g.\ suspensions of Brownian particles) is still
at its beginning \cite{szamel:2004}.

As a \emph{second} line of attack towards the realm far from
equilibrium, we want to mention the so--called \emph{non--equilibrium
work relations} or \emph{fluctuation theorems}.  These are theorems
supposed to remain valid beyond linear response, far from equilibrium,
and in particular also for small systems\footnote{Generally, care
needs to be taken that the notion of temperature is still valid, which
can be problematic for small systems \cite{gross:2004}.}, away from
the thermodynamic limit, where transient violations of the second law
of thermodynamics become increasingly likely. The fluctuation theorems
precisely say \emph{how likely}. The recent excitement about these
developments was initiated by the \emph{chaotic hypothesis} of
Gallavotti and Cohen \cite{gallavotti-cohen:95} and independent work
by Jarzynski \cite{jarzynski:98}, and it has even been suggested that
non-equilibrium statistical mechanics is presently undergoing a
foundational period comparable to that in equilibrium statistical
mechanics a century ago \cite{ruelle:2004}. The field has last been
reviewed in Refs.~\cite{evans-searles:2002,gallavotti:2002,%
maes-netocny:2003,crooks:2000,ritort:2003}. Here, we try to sketch the
basic idea. Consider a system prepared in thermal equilibrium with its
environment at temperature $T$ (e.~g.\ an RNA molecule). The system
shall be repeatedly subjected to an identical perturbation protocol
such that the external work $W$ is performed on it. In contrast to the
reversible work $W_0$, which is a state function, $W$ fluctuates from
one repetition to the next. If the average over the (infinitely many)
realizations of this process is denoted by an overbar and the
dissipative work by ${\cal W}\equiv W-W_0$, Jarzynski's equality may
be written in the form \cite{crooks:2000}
\begin{equation}\label{eq:jarzynski}
 \overline{e^{{\cal W}/k_BT}}= 1 \;.
\end{equation}
Obviously, the dissipative work must sometimes be positive as well as
\emph{negative}; i.~e.\ sometimes mechanical work will be gained from
the heat bath (see figure~\ref{fig:fluctuation_theorem}). This is
indeed what Einstein's theory of fluctuations \cite{einstein:10}
requires. Already if the protocol does not take the system too far
from equilibrium, so that fluctuations of ${\cal W}$ can be assumed to
be Gaussian distributed about an average dissipation ${\cal \overline
W}$, there will be some finite weight in the tail with negative ${\cal
W}$. This is not only consistent with the second law of
thermodynamics, the latter is in fact immediately deduced from
Eq.~(\ref{eq:jarzynski}) via Jensen's inequality ($\overline {e^x}
\geq e^{\bar x}$).  Equation~(\ref{eq:jarzynski}) can be used to infer
the reversible work $W_0$ from many repeated measurements of the
irreversible work $W$. This has been tested experimentally, e.~g.\ for
a Brownian particle dragged through water
\cite{wang-etal:2002,van_zon-cohen:2003}, and for RNA
\cite{liphardt-etal:2002} and a number of other biomolecules subjected
to mechanical forces (see Ref.~\cite{ritort:2003} for an overview and
suggestions for further experiments).

\section{Stochastic processes: the universal toolbox}
\label{sec:methods}

\subsection*{\bf Universality: the reign of large numbers}
In section~\ref{sec:history} we praised the elegance of Einstein's
derivation compared to Smoluchowski's attempt to construct a
microscopic dynamical theory of Brownian motion. This is not merely a
matter of elegance, though. While the strength of Smoluchowski's
approach lies in its ability to give insights into the microscopic
mechanism of Brownian motion, Einstein's approach relied on the idea
that certain aspects\footnote{Precisely which aspects, is currently
still an intriguing open question (see e.~g.\
\cite{baldassarri-colaiori-castellano:2003,ehrhardt-majumdar-bray:2004}).}
of fluctuations are independent of the details of the underlying
microscopic dynamic processes, as long as they are sufficiently many
and sufficiently uncorrelated in time and space (both aspects also
played a role in section~\ref{sec:gse}). If this is so, can one
develop a universally applicable calculus of stochastic phenomena?
Indeed, this is essentially what the theory of stochastic processes
aims at.

Einstein's derivation utilized the Chapman--Kolmogorov equation
Eq.~(\ref{eq:cke}) that translates the rules of a postulated
simplistic time--discrete jump pr
ocess of individual Brownian
particles encoded in the \emph{jump propabilities}
$\varphi_\tau(\xi)$, to the particle concentration
$n(x,t)$. Equivalently he could also have started from a formulation
that is discretized in space and continuous in time and write down a
\emph{master equation}
\begin{equation}\label{eq:meq}
\frac{d}{dt}P(x,t)=\sum_\xi[ w_{\xi\to x} P(\xi,t) - 
 w_{x\to\xi}P(x,t)]
\end{equation}
for a single particle with probabilities $P(x,t) \equiv n(x,t)/N$ and
\emph{jump rates} $w_{\xi\to x}$. Upon taking the limit of a large
number of independent jumps, one obtains along either way a
coarse--grained description in form of the diffusion equation
Eq.~(\ref{eq:deq}). It is important to note that in taking this limit,
some information about the underlying stochastic process, e.~g.\ about
the trajectory $x(t)$, is irretrievably lost. The Chapman--Kolmogorov
equation Eq.~(\ref{eq:cke}) and the master equation Eq.~(\ref{eq:meq})
provide a more detailed or microscopic description than the diffusion
equation, Eq.~(\ref{eq:deq}). Indeed, Einstein exploits this fact by
implicitely assuming that the difference between the actual
microscopic dynamics, which Smoluchowski struggled to calculate, and
the simplistic random walk will not matter for the end results
Eqs.~(\ref{eq:deq}),~(\ref{eq:msd}). The idea becomes maybe most
transparent from yet another equivalent approach Einstein could have
taken. Starting from the discrete trajectory
\begin{equation}\label{eq:rwt}
 x(t)=\sum_{n=1}^{N}\xi (n \tau)
\end{equation}
of a single particle with independent and identically distributed
random jump lengths $\xi$, he could have appealed to the law of large
numbers. Provided the first two moments of the jump distribution
$\varphi_\tau(\xi)$ exist, the cental limit theorem guarantees that
the limiting probability density for the individual particle
trajectories $x(t)$ starting at $t=0$ at the origin is given by the
Gaussian~\cite{feller:68}
\begin{equation}\label{eq:gf}
P(x,t)= \frac{1}{\sqrt{4 \pi D t}} \exp \left[ - \frac{x^2}{4Dt}
\right] \, .
\end{equation}
For the independent Brownian particles Einstein considered, this
statement is of course equivalent to saying that the
particle concentration $P(x,t)$ obeys the diffusion equation,
Eq.~(\ref{eq:deq}). In fact, Eq.~(\ref{eq:gf}) is nothing but the
Green function of Eq.~(\ref{eq:deq}), i.~e.\ its solution to the
initial condition $P(x,0)=\delta(x)$.

\begin{figure}[htb]
\begin{minipage}[t]{.45\textwidth}
\includegraphics[width=\textwidth]{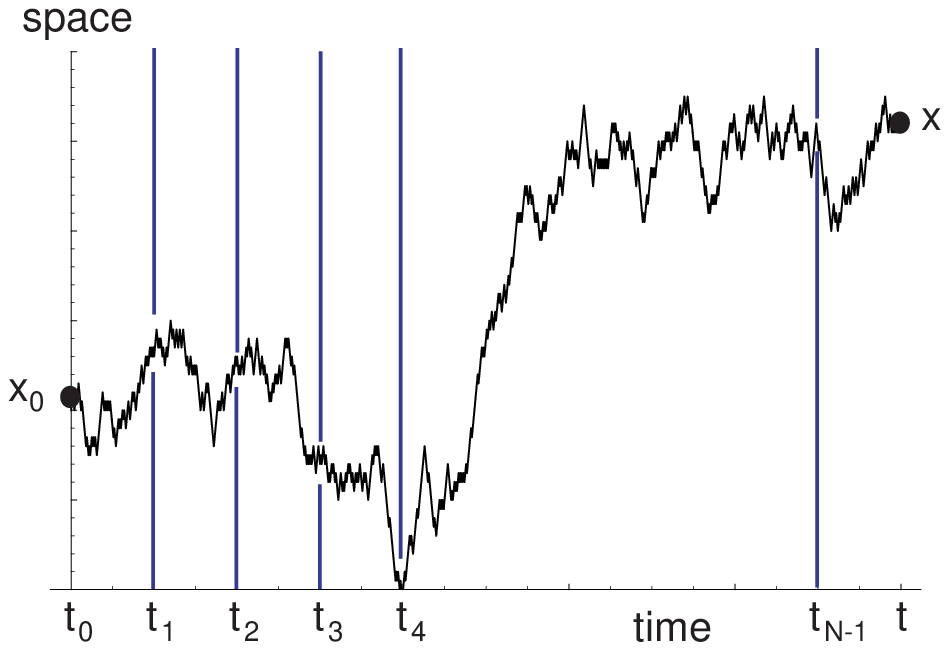}
\caption{Illustration of the concept of a path integral: 
  trajectory of a Brownian particle starting at position $x_0$ at time
  $t_0$ and passing through $N$ gates of size $d x_1, dx_2, \cdots,
  dx_{N-1}, dx$ at times $t_1, t_2, \cdots, t_{N-1}, t$.}
\label{fig:path_integral}
\end{minipage}
\hfil
\begin{minipage}[t]{.45\textwidth}
\includegraphics[width=\textwidth]{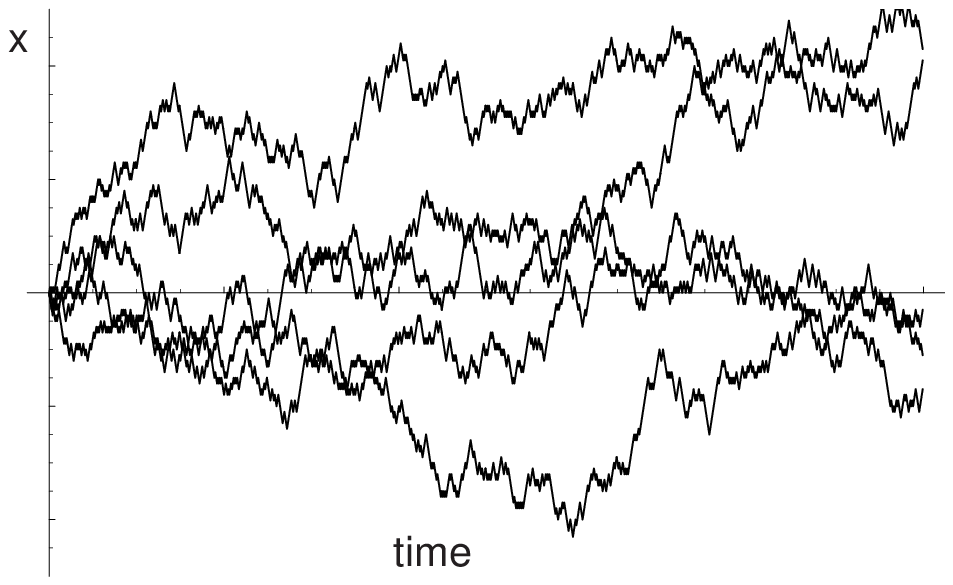}
\caption{Five different realizations of a random walk on a 
  one-dimensional lattice illustrating the fractal nature of the
  trajectories, which in the continuum limit are continuous but
  non-differentiable.}
\label{fig:fractal}
\end{minipage}
\end{figure}

Given such a powerful attractor of probability distributions, it is an
intriguing question, whether there is a particularly simple member of
the class of all stochastic processes with probability densities that
converge to Eq.~(\ref{eq:gf}). Is there a simplest representative
trajectory corresponding to the diffusion equation? This question was
positively answered by Wiener in 1921 \cite{wiener:21}. Inserting
Eq.~(\ref{eq:gf}) with a short time interval $t=\tau$ for the jump
probabilities $\varphi_\tau(\xi)$ in the Chapman--Kolmogorov equation
Eq.~(\ref{eq:cke}) and iterating the latter one arrives at the
expression
\begin{equation}\label{eq:wiener}
P(x,t)= \int \prod_{j=1}^{N-1} \frac{dx_j}{\sqrt{4 \pi D \tau}}
\; e^{- \sum_{j=0}^{N-1} \frac{(x_{j+1}-x_j)^2}{4 D \tau}} \to
\int_{x(0)=0}^{x(t)=x} \!\!\!\!{\cal D} [x(\tau)] \; e^{-\frac{1}{2D}
\int_0^t \frac 12\left( \frac{\partial x}{\partial \tau} \right)^2
d\tau}\;.
\end{equation}
The final \emph{path--integral} representation is obtained upon taking
the continuum limit $N\to\infty$ with $N\tau=t$, and has a
well--defined meaning within the framework of measure theory
\cite{wiener:21}. Wiener could also characterize the typical paths
corresponding to the limit taken in Eq.~(\ref{eq:wiener}) and show
that they are unphysical \emph{fractals}, continuous but
non--differentiable curves (see Fig.~\ref{fig:fractal}). One can
imagine them to be constructed by a self--similar continuation of the
large--scale structure of a trajectory of a Brownian particle to
arbitrarily short distances. The intricate details of the actual
physical process at short scales together with the corresponding
characteristic microscopic length scale are simply wiped off to give
way to the ultimate simplicity of ideal self--similarity. The innocent
looking prescription given after Eq.~(\ref{eq:wiener}) actually
encodes a \emph{renormalization} of the original discrete stochastic
model to a mathematically more convenient scale--free phenomenological
model, known as the \emph{Wiener process}.

\subsection*{\bf The art of mapping: from Brownian motion to quantum
  mechanics and back}

{F}rom what was just said, it is natural to expect that the Wiener
process finds applications in many different areas far beyond the
proper domain of Brownian motion, which is indeed the case
\cite{kleinert:2004}. Here we restrict ourselves to a few examples.
Among physicists even better known than the Wiener measure
Eq.~(\ref{eq:wiener}) is the Feynman path integral
\cite{feynman-hibbs:65}, which expresses the propagator or Green
function of a non--relativistic free particle in quantum mechanics as
a path integral as in Eq.~(\ref{eq:wiener}) but with an imaginary
exponent. More precisely, the prefactor $-1/2D$ is replaced by
$i/\hbar$.  Within the path integral picture one can interprete the
uncertainty relation of quantum mechanics as the analog of the
diverging apparent velocity of the Brownian particles that worried
experimentalists before 1905. Note that the exponent in
Eq.~(\ref{eq:wiener}) has the form of the Lagrangian of a classical
free particle. This analogy can be extended \cite{kac:49} to include a
potential energy $-U(x)$ under the integral in the exponent of the
Wiener measure. The corresponding Feynman path integral (the
Feynman--Kac formula) can intuitively be understood as an analytic
continuation of this extended Wiener measure in the complex plane, but
a proper mathematical formulation is more subtle \cite{kleinert:2004}.
The diffusion (or \emph{Fokker--Planck}) equation corresponding to the
extended Wiener measure is Eq.~(\ref{eq:deq}) extended by a term
$-U(x) n(x,t)$, corresponding to an annihilation probability.  The
diffusion equation corresponding to the Feynman--Kac propagator is of
course the \emph{Schr\"odinger equation} (in imaginary time).

Well known to physicists, this workhorse of quantum mechanics was the
starting point for the development of various analogies. Building on
the pioneering works, many important applications of path integrals
have been developed in statistical physics and quantum and classical
field theories, as well as in polymer physics and finance (where
quantum mechanics was essentially mapped back onto Brownian motion),
each field contributing its own characteristic difficulties to the
basic formulation. Out of the innocent jittering of a pollen grain has
grown a universal and powerful tool of theoretical physics, which has
helped to unravel intimate connections between what seem to be vastly
different phenomena at first sight.

\begin{figure}[htb]
\begin{center}
\includegraphics[width=0.8\textwidth]{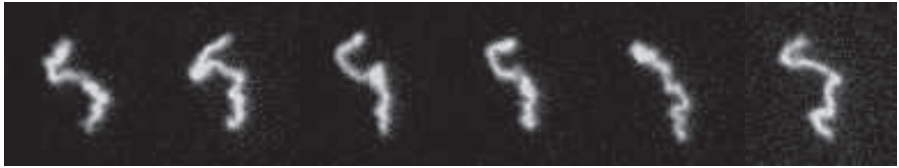}
\end{center}
\caption{Snapshots of fluorescently labeled DNA
  molecules adhering to a two--dimensional lipid membrane. Courtesy of
  J. R{\"a}dler and B. Maier~\cite{maier_raedler:99}.}
\label{fig:DNA}
\end{figure}

We briefly sketch some key developments in the application to static
conformations of flexible polymers. If segments of a flexible polymer
could freely pass through each other, the Wiener process (in this
context known as the \emph{Gaussian chain}) would indeed provide an
excellent coarse--grained description of the conformation if time $t$
is identified with arclength $s$ so that $6Dt$ is replaced by the
mean--square end--to--end distance $\avg{R^2}$ of the
polymer. In particular, the Stokes--Einstein relation is translated
into the ``rubber'' force--extension relation  
\begin{equation}\label{eq:rubber}
 \frac{\partial \avg{R}_F}{\partial F} = \frac{\avg{\ve R^2}}{3k_BT} \;,
\end{equation}
which is easily recognized as the soft--matter analog of the Curie law
for the inverse--temperature susceptibility of a paramagnet. Both
become ``stiffer'' with increasing temperature, due to the
disarranging action of thermal fluctuations. However, except under
so--called $\theta-$conditions, where the self--avoidance of the chain
is screened by an effective self--attraction, the simple random walk
is unrealistic because of this idealization. Real polymers
conformations cannot intersect and are characterized by
\emph{self--avoiding} random walks.  The {\em Edwards model}
\cite{degennes:72,doi-edwards:86} represents self--avoidance by
substituting the potential in the extended Wiener measure by a
singular repulsion $U(\ve r) \propto \delta[\ve r (s)-\ve r
(s')]$. This model has been the starting point of much theoretical
work in polymer physics~\cite{descloizeaux:90} and has been
generalized to other self-avoiding
manifolds~\cite{membrane:winterschool}. It is amenable to a rigorous
treatment with perturbation series and renormalization methods based
on the path integral~\cite{descloizeaux:90,schaefer:99}. Notably, as
has been pointed out by de Gennes~\cite{degennes:72}, the problem of a
self--avoiding random walk is related to critical phenomena, as it can
also be mapped onto the $n\to0$ limit of a $O(n)-$field theory.
Figure~\ref{fig:DNA} shows snapshots of fluorescently labeled DNA
molecules adhering to a two--dimensional lipid
membrane~\cite{maier_raedler:99}. Measurements of the radius of
gyration $R_g$ of the polymer have very well confirmed the prediction
$R_g\propto L^{3/4}$ for the (``super--diffusive'') conformation of a
self--avoiding random walk in two dimensions. However, the interest of
physicists for DNA is by no means limited to such very coarse and
generic properties. Over the last 10 years DNA has become a favorite
model system for a whole new field of investigation one might call
``bio--polymer physics''
\cite{frey-etal:97,bustamante-bryant-smith:2003,peyrard:2004,nelson_david:2004}.

Self--similar renormalized models are mathematically convenient but
fail to indicate where the simplified universal representation breaks
down.  If one needs to incorporate more details, one has to resort to
the level of the Chapman--Kolmogorov equation, Eq.~(\ref{eq:cke}), or
master equation, Eq.~(\ref{eq:meq}). The latter has many features in
common with the second quantized Schr\"odinger equation for
many--particle quantum systems. In particular, both are linear
differential equations of first order in time. Since each of the
possible states $\alpha$ of the system can be labeled by a set of
occupation numbers $n(x,t)\to\{ n_i(t) \}$ (e.~g.\ of lattice sites,
energy states, chemical states, etc.), one is immediately led to a
formulation in terms of an operator algebra in analogy with the second
quantized formulation of quantum mechanics (for a short review and a
list of references see e.~g.~\cite{taeuber:03}).  Upon introducing
boson ladder operators with the standard commutation relation $[a_i,
a_j^\dagger] = \delta_{ij}$, one can define Fock states $| \alpha
\rangle = \prod_i (a_i^\dagger)^{n_i} | 0 \rangle$ via acting with the
ladder operators on the ``vacuum'' state $| 0 \rangle$. Then the
master equation for the probability $P(\{n_i\},t)$ of finding the
system in the state with $\{ n_i\}$ particles on site $\{i\}$,
respectively, is equivalent to a Schr\"odinger equation in imaginary
time
\begin{eqnarray}
  \label{eq:schroedinger}
  \frac{d}{dt} | \Psi \rangle = - {\cal H} | \Psi \rangle
\end{eqnarray}
for the state vector $|\Psi \rangle = \sum_\alpha P(\{n_i\},t) |\alpha
\rangle$. Of course, there are also differences from quantum
mechanics. Most importantly, the ``Hamiltonian'' $\cal H$ is in
general non--Hermitian. Still, most if not all many body techniques
can be applied. In particluar, it is possible to transform the second
quantized Hamiltonian into a field theory upon using coherent-state
path integrals~\cite{taeuber:notes}.

The similarities of the independently developed formalisms have led to
a successful description of classical statistical systems far from
equilibrium~\cite{taeuber:03}. They not only allow for an application
of field theoretical methods but also for a fruitful exchange of
results and concepts between seemingly unrelated phenomena in quantum
and classical physics. Very recently, it was e.~g.\ suggested to
represent gene expression as a many--body quantum
problem~\cite{sasai_wolynes:03}, which leads to such unexpected
analogies as that between the stochastic dynamics of a single gene
switch and the quantum mechanics of the spin--boson model of a
two--site polaron.

\section{A noisy world: the ubiquity of stochastic processes
  in biology}
\label{sec:noisy_world}

\subsection*{\bf Genetic drift and the theory of neutral evolution}
Today we live in a ``noisy world'' where it is difficult to find a
quiet place to contemplate. Noise in this context is simply unwanted
sound and certainly deleterious.  However, at the very origin of our
lives are random processes, i.~e.\ noise. Luckily, one should say,
because otherwise we would largely be deprived of our individuality
and doomed to live in a world of clones. This insight we owe to
Mendel, who deduced from his experiments with pea flowers the
existence of genes present as pairs of alternative forms (called
alleles) in each (somatic) cell.  He assumed that reproduction happens
in a two step process, where equal separation of these gene pairs into
the gametes (egg and sperm) is followed by {\em random fertilization},
i.~e.\ random union of one gamete from each parent to form the first
cell (zygote) of a new progeny's individual. This process is random
precisely because there is Brownian motion on a molecular scale.

\begin{figure}[htb]
\begin{minipage}[t]{.45\textwidth}
\centerline{\includegraphics[width=0.7\textwidth]{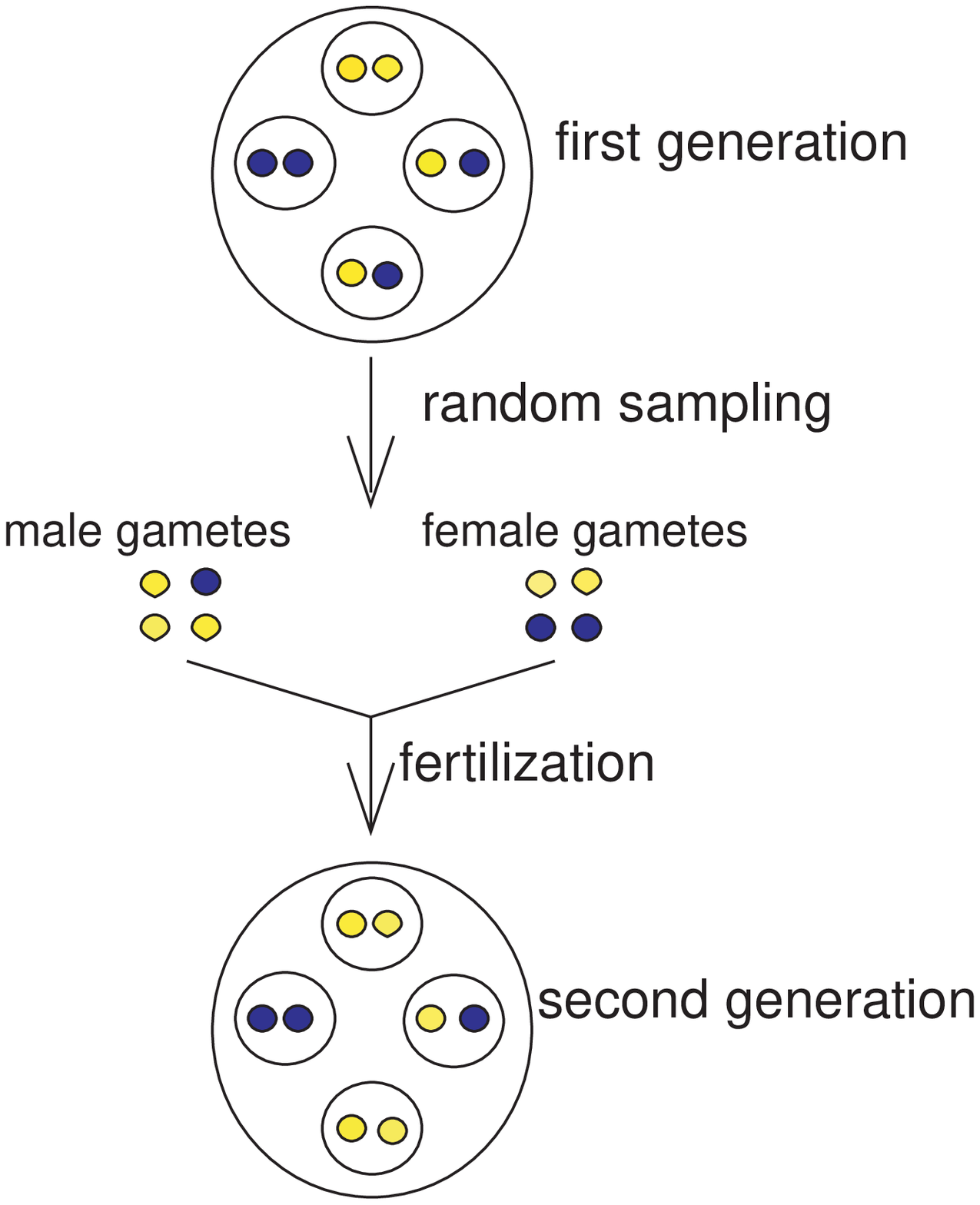}}
\caption{Panmixia: a toy model of population genetics.}
\label{fig:panmixia}
\end{minipage}
\hfil
\begin{minipage}[t]{.45\textwidth}
  \includegraphics[width=\textwidth]{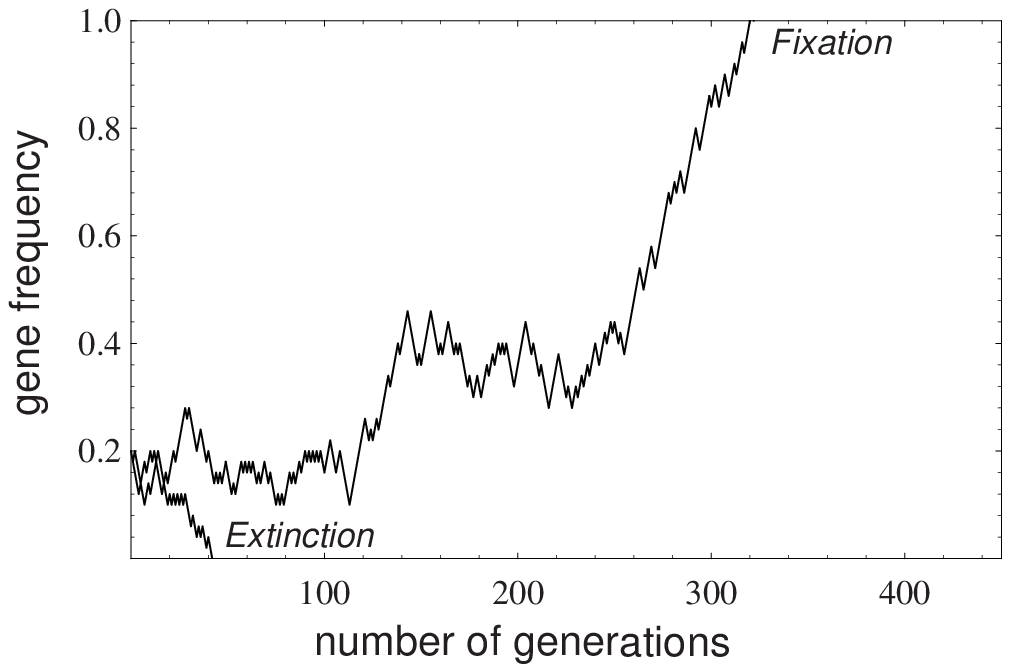}
\caption{Genetic drift. Some representative sample paths for
  random gene frequency drift.}
\label{fig:genetic_drift}
\end{minipage}
\end{figure}

Starting at about the same time as Einstein's fundamental paper on
Brownian motion the re-discovery of the Mendelian laws prompted the
emergence of a new discipline, now known as {\em population genetics}
(for an historic account see e.~g.\  Ref.~\cite{kimura:83}). Pretty
early on it was realized that random fertilization in finite
populations leads to a phenomenon called {\em ``genetic
  drift''}~\cite{fisher:22}.  To illustrate this point let us look at
the following toy model of population genetics (see
Fig.~\ref{fig:panmixia}), where one considers only one gene locus in a
population of diploid\footnote{The genome is made up of one or more
  extremely long molecules of DNA that are organized into chromosomes.
  The body cells of most plants and animals contain two genomes. These
  organisms are called diploid. For more details the reader is
  referred to a recent edition of some textbook in genetics, e.~g.\
  \cite{griffiths_etal:book}.}  organisms that has two alleles $A$ and
$a$. We assume that each individual releases the same number of
gametes, so that the allelic frequency in the population of gametes
and the parental population are identical. From this large pool of
gametes the next progeny is obtained by randomly chosing $N$ pairs of
gametes (``random mating'' or ``panmixia''). In the language of
sochastic processes this is a sequence of $2 N$ independent {\em
  Bernoulli trials}.  If we denote by $x$ the relative frequency
(probability) for the allele $A$ in the parental population, the
probability distribution for the number $n$ of alleles $A$ in the next
generation is given by the {\em binomial distribution}
\begin{eqnarray}
  B(n,N) = \frac{N!}{n!(N-n)!} x^n (1-x)^{N-n} \, .
\end{eqnarray}
Therefore, even if there are no deterministic evolutionary forces
acting on a population (such as natural selection), sampling errors
during the production of zygotes may cause a change in the frequency
on a particular allele in a population; this phenomenon is called
genetic drift. As can be read off from the variance of the binomial
distribution, $D (x) = x(1-x)/2N$, genetic drift is expected to be
large in small populations and vanishes for infinite population size.
In the language of Brownian motion ``genetic drift'' is actually no
drift at all, but a random walk of the stochastic variable $x$ (gene
frequency). Two examples of such walks either leading to extinction or
fixation are shown in Fig.~\ref{fig:genetic_drift}.

In complete analogy to Einstein's formulation of Brownian motion one
can now consider an ensemble of populations and derive a differential
equation for the conditional probability density $n(x,p;t)$ that the
gene frequencs lies between $x$ and $x + dx$ at time $t$, given that
the initial gene frequency was $p$. It is interesting to note that
such a formulation was given by Fisher~\cite{fisher:22} without being
aware of parallel developments in physics. The corresponding
Fokker-Planck equation reads
\begin{eqnarray}
  \frac{\partial n (x,p;t)}{\partial t} = \frac12
  \frac{\partial^2}{\partial x^2} 
  \left[ 
  D (x) \, n (x,p;t)
  \right] \;  ; \quad \text{with} \quad D(x) = x(1-x)/2N \, ,
\label{eq:fpe_genetic_drift}
\end{eqnarray}
which can be interpreted as a diffusion process with progressively
slower motion near the boundaries of gene frequency space, i.~e.\ $x=0$
and $x=1$. Similar models with spatially varying diffusion
coefficients have more recently been studied in the context of the
voter model~\cite{liggett:book,krapivsky_redner:96} and of some
catalysis reaction models at surfaces~\cite{Ben-Avraham_etal:90}.

An essential feature of the random ``drift'' in gene frequency space
is that there are boundaries. Once a given allele $A$ reaches $x=0$ it
becomes extinct and the random process stops.  Similarily if the
allele $A$ reaches a frequency of $x=1$ it is said to be fixed (or
equivalently allele $a$ becomes extinct). In other words, what we are
facing here is a diffusion process with two {\em absorbing
  boundaries}. One may now ask the following questions. What is the
time evolution for the distribution of unfixed classes with $0<x<1$?
How likely is it for an allele to become fixed and what is the
statistics of the number of generations until extinction or fixation
of a certain allele occurs? Both of them can be answered by analyzing
the Fokker-Planck equation Eq.~\ref{eq:fpe_genetic_drift}, whose
application to population genetics has been pioneered by Kimura, one
of the founding fathers of the {\em neutral theory of
  evolution}~\footnote{The neutral theory of evolution asserts that
  the great majority of evolutionary changes at the molecular level
  are caused not by Darwinian selecion but by random drift of
  selectively neutral mutations~\cite{kimura:83}.}. He
showed~\cite{kimura:55} that in the asympotic limit of a large number
of generations the distribution function of the unfixed classes
becomes flat and decays exponentially as $n (x,p;t) = 6 x (1-x) \exp
(-t/2N)$.  For answering the second question there is an alternative
route using concepts from game theory.  To see this let us shortly
digress to a phenomenon called {\em ``gambler's ruin''}.  Say the game
simply consists of repeatedly throwing a fair coin. Your initial
capital is $p$ (with $0<p<1$) and that of your opponent is $1-p$. The
game is over when either one of you goes broke.  What are your chances
to win the game?  The answer is simple (see e.~g.\
Ref.~\cite{jones_smith:book}). Your chances are $p$, i.~e.\
proportional to your initial capital, which can be rather low if you
try to play against someone with large reserves\footnote{But history
  teaches us that even the big players with large reserves may
  sometimes stumble and fall \cite{dunbar:2000}.}.  This is exactly
the game played by the genes in the above toy model of population
genetics. Hence the probability that the allele $A$ eventually becomes
fixed (``wins'') is identical to its initial frequency $p$. But how
long will it take to win? This is a first passage time problem (for a
review see e.~g.\  Ref.~\cite{redner:book}) and can be solved by using a
cousin of the Fokker-Planck equation, the backward Kolmogorov
equation~\cite{kampen:92}. If one starts with just a single
``invading'' allele $A$ with a frequency of $p=1/2N$ in a population
of size $N$, one finds~\cite{kimura_ohta:69} that both the mean and
the standard deviation of the time to fixation grow linearly with the
size of the population.

This simple toy model of population genetics with one gene locus, two
alleles and random mating is, of course, much to naive to account for
the variety and complexity of life forms on our planet. On a molecular
level, there are complex changes in DNA sequence, which may either be
random-like point mutations and frameshifts or more systematic
cut-and-paste mechanisms like insertion sequences and
transposons~\cite{kazazian:04}. On the phenotype level, there are
interactions among the organisms in a population and with the
surrounding environment, which may itself be dynamic and noisy. The
fitness, i.~e.\ the number of expected offsprings, of an individual is a
(highly non-linear) function of such interactions. The way from
genotype to a phenotype is paved by networks of coupled biochemical
reactions and feedback signals. The list goes on and would lead us
considerable off track; for more information the interested reader is
recommended to consult some recent introductory review
articles~\cite{drossel:01,baake_gabriel:00}.

\subsection*{\bf  Stochasticity in chemical reactions: the reign of
  small numbers} 

The machinery of biological cells consists of networks of molecules
interacting with each other in a complex manner. Many of these
interactions can be described as chemical reactions, where the
intricate processes which occur during the encounter of two molecules
are reduced to reaction rates\footnote{Reaction rate theory was
  pioneering by Kramers~\cite{kramers:40}; for a recent review see
  e.~g.\ \cite{haenggi_talkner_borkovec:90}. He considered a simplified
  model system where a Brownian particle is trapped in a
  one-dimensional well representing the reactant state which is
  separated by a barrier of finite height from a deeper well
  signifying the product state.}.  For example, an {\em autocatalytic
  reaction} for an enzyme $E$ is modeled as a probabilistic
self-replication process $E \to 2 E$, which occurs at a constant
probability per unit time (reaction rate) $k$.  For high
concentrations of the enzyme, a standard approach in chemistry is to
write down {\em rate equations} for the time evolution of the average
concentration $\bar n (t)$
\begin{eqnarray}
  \frac{d \bar n (t)}{dt}  = k \, \bar n (t) \, .
\label{eq:rate_autocatalytic}
\end{eqnarray}
Such an approach evidently assumes that the time evolution of a
chemically reacting system is both {\em deterministic} and {\em
  continuous}.  However, the actual process is neither of both since
molecular population levels change {\em stochastically} and by {\em
  discrete} integer amounts. 

Obviously, a deterministic description is applicable only if
deviations from the average number of molecules are negligibly small.
As noted long ago by Delbr\"uck~\cite{delbrueck:40} the validity of
this assumption breaks down in small chemical systems, where the
concentrations of reacting species are extremly low and the
statistical fluctuations of atomism can not be avoided. Then, instead
of the average concentration $\bar n$ one now has to consider the full
probability distribution function $P_n (t)$ of the number of enzymes
$n$ at a given time $t$. This function plays the same role as the
particle density $n(x,t)$ in Einstein's treatment of Brownian motion
or the probability density of gene frequency $n(x,p,t)$ in our
discussion of genetic drift. Instead of continuous random variables
(particle position, gene frequency) we are now dealing with the
discrete random variable $n(t)$, the number of enzymes at time $t$.
The time evolution for $P_n (t)$ can immediately be written down if
one pictures an ensemble of enzymatic reactions, where $P_n$ is the
relative frequency of samples in the ensemble with exactly $n$
molecules. The frequency $P_n$ (population size) of this particular
class is reduced by each self-replication process within the class and
enhanced by each self-replication processes in the neighboring class.
Then the time evolution of $P_n (t)$ can be written as a ``gain-loss''
equation
\begin{eqnarray}
    \frac{d}{dt} P_n (t) = (n-1) k P_{n-1} (t) - nk P_n (t) \, ,
\label{eq:master_autocatalytic}
\end{eqnarray}
a specific example of the {\em master equation} introduced in
section~\ref{sec:methods}.

How are the results of a stochastic description,
Eq.~\ref{eq:master_autocatalytic}, different from a deterministic
approach, Eq.~\ref{eq:rate_autocatalytic}? The most striking one is,
of course, that despite the exponential growth, $n(t)=n_0 \exp(kt)$,
promised by a deterministic description there is a chance to loose it
all, $n(T)=0$ (gambler's ruin). Other features of the stochasticity
become evident upon solving the master equation for reactions
initiated by one particle. Then the probability distribution for large
times $t$ has the asymptotic form
\begin{eqnarray}
  P_n^1 (t) = \frac{1}{\bar n} \, e^{-n/\bar n} 
  \quad {\rm with} \;\; 
  \bar n = e^{kt} \, ,
\label{eq:limiting_auto}
\end{eqnarray}
which is by no means normal, but exponential.  Note that at all times
the probability of finding a single particle is greater than the
probability of finding any other specified number of particles. In
addition, the standard deviation is idential to the mean of the
distribution. Delbr\"uck also points out that there are strong
fluctuations in the waiting time for attaining a certain threshold
amount of reaction products~\cite{delbrueck:40}.

An even more dramatic illustration for the importance of stochasticity
in chemical reactions is given by the pair annihilation process $A + A
\to 0$ in one-dimensional systems. The rate equation, $dn / dt = - k
n^2$ would predict an algebraic decay as $n(t) \sim t^{-1}$. However,
the actual asymptotic decay is much slower, $n(t) \sim t^{-1/2}$. This
slow decay can be traced back to the reentrance property of random
walks in less than two dimensions, which implies that within a volume
of order $\sqrt{Dt}$ all particles are annihilated with probability
one~\cite{toussaint_wilczek:83}. A more sophisticeted line of argument
uses methods adapted from quantum field theory~\cite{lee94}. The pair
annihilation process is the simplest member of a broad class of models
known as diffusion-reaction models exhibiting absorbing states,
recently reviewed in Ref.~\cite{hinrichsen:00}. That such systems show
unusual dynamics becomes obvious by asking the innocent looking
question: How does a system evolve towards its steady state? Sometimes
the answer is quite simple and the relaxation process is merely an
exponential decay. If the deviations from equilibrium are small
Onsager's regression hypothesis~\cite{onsager31} asserts that the
relaxation is governed by the same laws as the fluctuations in
equilibrium. But this hypothesis certainly fails for systems with an
absorbing steady state.  Here there are no fluctuations in the steady
state but the approach towards the absorbing state is critical in the
sense that it exhibits slow power-law decay and universal scaling
behavior~\cite{cardyreview}.

Before getting too much sidetracked let us come back to biology.  Here
it has been emphasized recently that living cells possess very low
copy numbers of many components, including DNA and important
regulatory molecules~\cite{guptasarma:95}. As a consequence,
stochastic effects such as those illustrated above may play a key role
in biological processes. A quite striking example are
observations~\cite{spudich_koshland:76,mcadams_arkin:97} that clonal
populations of cells (i.~e.\  genetically identical cells) exhibit
substantial phenotypic variation. Recent experiments even allowed to
distinguish between the stochastic noise inherent in the chemical
reactions controlling gene regulation and fluctuations in the number
of other cellular components~\cite{elowitz_etal:02}. These
observations are only the tip of an iceberg. ``Noisy'' mechanisms may
well be rather the rule than the exception in the molecular world of a
cell~\cite{mcadams_arkin:99}.

\section{The Brownian particle: probe and effigy of the intangible}
\label{sec:probe}

\subsection*{\bf Probing molecular fluctuations with Brownian 
  particles: ``the buoy in the surge''} 

Precisely in the way proposed by Einstein \cite{einstein:05}, Brownian
particles have ever since been used as probes for the molecular motion
of the surrounding medium: recently in particular for probing the
local viscoelastic response of soft, often biological, materials,
where elastic moduli are of the order of $k_BT$ or even due to $k_BT$
(i.~e.\ of entropic origin).  A zoo of methods based on Brownian motion
of tracer particles has been invented: (\emph i)
\emph{particle--tracking} techniques\footnote{Roughly speaking, even
the latest computer--aided confocal video--microscopy setup for
particle tracking is but a costly late descendant of the technique
pioneered by Henri in 1908 \cite{henri:08}, when he first put a film
camera onto a microscope to record the wild paths of Brownian
particles in order to scrutinize Einstein's prediction,
Eq.~(\ref{eq:msd}).} that record the path of one particle
\cite{denk-watt:89,amblard-etal:96,gittes-etal:97} or the correlated
fluctuations of two particles
\cite{crocker-etal:2000,forstner-etal:2003}; (\emph{ii})
\emph{diffusing light spectroscopy}
\cite{maret-wolf:87,weitz-pine:93}, which iterates our basic theme by
exploiting not only the diffusion of colloids but also the
\emph{diffusion of light} they scatter; (\emph{iii})
\emph{Fourier--imaging--} \cite{schurr:2000} and \emph{fluorescence--}
\cite{rigler-elson:2001} \emph{correlation--spectroscopy}, which are
more similar to classical dynamic scattering techniques; etc.
According to the fundamental connection between fluctuations and
response provided by Eqs.~(\ref{eq:ser}),
(\ref{eq:msd}),(\ref{eq:gkr}) the fluctuations measured by these
\emph{passive} techniques contain the same information as obtained by
\emph{active} microrheological linear--response measurements (e.~g.\
\cite{bausch-moller-sackmann:99}). One can roughly classify the
passive methods into those working in Fourier space (``scattering'')
and those working in real space (``microscopy'').

The basic principle of scattering methods probing Brownian dynamics is
well illustrated by the example of \emph{particle sizing}, which is a
widely applied standard method. One shines a laser beam of wavelength
$\lambda$ at a dilute suspension of $N$ identical particles in a
viscous medium and records the light intensity scattered at an angle
$\theta$ to the incident beam. The auto--correlations of the scattered
light then decay proportional to (the square of) the dynamic structure
factor
\begin{equation}\label{eq:sqt}
S_q(t)=\frac1N\sum_{ij}\avg{e^{i\ve q \cdot [\ve r_i(t)-\ve r_j(0)]}}
\stackrel{\rm dilute}{\sim}\frac1N\sum_{i}\avg{e^{i\ve q \cdot
\delta\ve r_i(t)}}= e^{-q^2\avg{\delta \ve r_i(t)^2}/6}= e^{-q^2Dt}\;.
\end{equation}
Here, $\ve q$ is the scattering vector, $q\equiv |\ve q| \equiv
4\pi{\sf n}\sin(\theta/2)/\lambda$ with $\sf n$ the refractive index,
and we used that dilute particles are uncorrelated.  Applying
Einstein's Eq.~(\ref{eq:msd}) together with the Stokes relation
$\zeta=6\pi\eta a$ for the friction of a spherical particles in a
viscous solvent, one can determine the diameter $a$ of the particle or
alternatively the vicosity $\eta$ of the solvent.

\subsection*{\bf Brownian particles in complex fluids: 
  ``the buoy in the pudding''} 

Although the idea of exploiting such kind of methodology to measure
material properties in small samples and even inside living cells is
also a rather old one \cite{heilbronn:22}, its applications mulitplied
over the last two decades or so with surgeing interest in highly
disordered, soft, and/or biological matter. Reports of substantial
deviations from Eq.~(\ref{eq:msd}) in such systems abound. A common
observation is that of \emph{superdiffusive} ($\alpha>1$) or
\emph{subdiffusive} ($\alpha<1$) dynamics, which was studied
theoretically in great detail \cite{bouchaud-georges:90}, and which is
characterized by
\begin{equation}\label{eq:sbd}
\avg{\delta \ve r^2}\propto t^{\alpha} \qquad \text{with } \alpha\neq 1 \;.
\end{equation}
Such ``strange kinetics'' \cite{shlesinger-zaslavsky-klafter:93} has
even prompted the development of a dedicated ``fractional calculus''
\cite{mandelbrot:83,tsallis-etal:95,metzler-klafter:2000}. It appears
in many areas of science (nonlinear dynamics, growth phenomena,
fluctuating manifolds, etc.). A simple way to generate such anomalous
diffusion is by assuming jump rates $\varphi_\tau(\xi)$ in
Eqs.~(\ref{eq:cke})~(\ref{eq:rwt}) that decay too slowly with $\xi$
(i.~e.\ like $\xi^{-\mu}$, $1<\mu<3$) for the ordinary central limit
theorem to apply, or, equivalently, for constant step length by a
broad waiting time distribution. For Brownian particles, it naturally
arises as intermediate asymptotics whenever the solvent is not a
simple liquid but a complex (viscoelastic) fluid with some internal
structure that exhibits long--range spatial or temporal correlations;
i.~e.\ it has more to do with a complicated material property of the
solvent than with Brownian motion as such, and thus leads us slightly
off--topic. Biological cells certainly provide us with ultimate
complexity, which makes deviations from Eqs.~(\ref{eq:msd}),
(\ref{eq:sqt}) anything but surprising. It is therefore very
suggestive to interprete subdiffusion in cells as a consequence of
cytoplasmic heterogeneities and crowding
\cite{weiss-etal:2004,tolic-noerrelykke:2004}, but attempts to explore
the actual connection are still limited to model systems
\cite{ott_etal:90,wong-etal:2004}.  If not due to complex (biological)
organization, one generally expects long--range correlations to be
\emph{self--similar}, arising e.~g.\ as a consequence of a closeby
equilibrium critical point, turbulence, or some other non-equilibrium
structure. A nice example of temporal correlations is provided by
simple shear flow: it gives rise to a peculiar cooperation of
statistical mechanics and fluid mechanics known as Taylor diffusion
\cite{ben_naim-redner-ben_avraham:92,mazo:2002}, which effectively
speeds up the diffusion of (dilute) suspended Brownian particles.
Stirring your coffee thus helps twice to distribute the suger, via
turbulent convection on large scales and via the induced
superdiffusion on shorter scales. A common paradigm based on spatial
correlations is diffusion in a porous medium with a very broad pore
size distribution, idealized as diffusion on a fractal
\cite{benavraham-havlin:2000,stauffer-aharony:94}. More intriguing
from a theoretical perspective is the case, where subdiffusion arises
without any non--trivial structures in configuration space as a
consequence of a complicated phase--space structure. This kind of
``dynamically self--generated'' subdiffusion is often characterized as
\emph{glassy dynamics} (see section~\ref{sec:gse} and below).

\subsection*{\bf Long--time tails: ``the buoy's bow wave''}

Maybe most suprisingly, already the dynamics of a single Brownian
particle in a simple Newtonian solvent has some subtle properties not
anticipated by Einstein and Perrin. In his derivation, sketched above,
Einstein implicitely assumed $\varphi_\tau(\xi)$ to have properties
such that the expansion of the Chapman--Kolmogorov equation,
Eq.~(\ref{eq:cke}), nowadays known as \emph{Kramers--Moyal expansion},
can be truncated after the $\nabla^2-$term, thus yielding a simple
hydrodynamic equation, the diffusion equation, Eq.~(\ref{eq:deq}).
Such conditions are mathematically realizable
\cite{nelson:67,gardiner:85,tzanakis-grecos:99} but not necessarily
physically relevant \cite{mclennan:88,kampen:92}. They amount to a
description of Brownian motion as a \emph{Gaussian Markov process} (a
notion that did not yet exist in 1905). Although this may seem to be a
very plausible assumption given the law of large numbers and the
strong scale separation between the Brownian particle and the
molecular scale, it only holds to a first approximation, or for
$t\to\infty$. Hydrodynamic conservation laws, as e.~g.\ contained in
the Stokes equation for the solvent, induce long--range
auto--correlations that effectively retard the approach to the Markov
limit. At any finite time, they contribute transient corrections,
so--called \emph{long--time tails}, to Eq.~(\ref{eq:deq}) and moreover
render Eq.~(\ref{eq:cke}) and the subsequent expansion ill defined
\cite{dorfman-cohen:65,alder-wainwright:67,ernst-hauge-van_leeuwen:70}
(for overviews see e.~g.\ \cite{keyes-masters:85,mclennan:88} and
references therein). A quick derivation exploits
Eq.~(\ref{eq:gkr}). Recalling that the dynamics of the solvent is
itself governed by a diffusion equation (see footnote 2,
section~\ref{sec:history}, above) we conclude that the momentum of the
Brownian particle is diffusively dissipated, hence shared with a
growing solvent volume $\Delta(t)^3\propto (\nu t)^{3/2}$, so that the
velocity auto--corelations in Eq.~(\ref{eq:gkr}) decay like
$t^{-3/2}$.  The effective diffusion coefficient $\avg{\delta\ve
r^2(t)}/6t$ at finite time $t$ is obtained by setting the upper limit
of integration in Eq.~(\ref{eq:gkr}) to $t$, which immediately gives
\cite{landau-lifshitz:fm,mclennan:88}
\begin{equation}\label{eq:ltt}
\avg{\delta\ve r^2(t)}=6Dt\left[1-\sqrt{\theta/t} + \dots\right]\;,
\end{equation}
where $\theta$ is a hydrodynamic time scale that depends on $\zeta$
and the particle/solvent mass ratio.  Thus, normal Brownian motion
actually is slightly faster than diffusive, Einstein's formula is only
slowly approached, and there is no simple general way to ammend the
diffusion equation to take this into account.  The somewhat
uncomfortable picture \cite{mclennan:88}, which first emerged from
calculations for dense gases based on the Boltzmann equation
\cite{dorfman-cohen:65}, has sometimes been summarized as Dorfman's
lemma \cite{evans-morriss:90}: \emph{all relevant fluxes are
nonanalytic functions of all relevant variables}. Neither does a
gradient expansion of the diffusion current exist (the
Burnett-coefficients appearing in the expansionn are found to be
infinite), nor do density expansions exist of the diffusion
coefficient and other transport coefficients as one might expect in
analogy with the virial expansion of thermodynamics. Luckily, the
corrections in Eq.~(\ref{eq:ltt}), though of fundamental interest, are
in general numerically too small to spoil seriously any of the
numerous applications mentioned above. We note in passing that related
problems arise in statistical mechanics derivations \emph{\`a la}
Einstein and Smoluchowski of hydrodynamic boundary conditions (e.~g.\
``no--influx'' for a tagged Brownian particle in a colloidal
suspension \cite{fuchs-kroy:2002} or ``no--slip'' in a Newtonian fluid
\cite{wolynes:76,bocquet-barrat:94,hagen-etal:97}). Again, typical
microrheological applications may well get by without such subtleness
\cite{gardel-etal:2003}. However, one should be aware that tails do
seriously affect Brownian motion in lower dimensions, leading to
``hydrodynamic suicide'' \cite{brenig:89} in two dimensions. Even
\emph{without} considering solvent hydrodynamics, due to the otherwise
subdominant long--range concentration patterns (``colloidal tails'')
implied by Eq.~(\ref{eq:deq}), Brownian motion in 1--dimensional pores
exhibits anomalous, subdiffusive dynamics, so--called single--file
diffusion \cite{lutz-kollmann-bechinger:2004}.

\subsection*{\bf Interacting Brownian particles: the upscaled microcosmos}

Apart from providing us with a large assortment of colloidal particles
to probe the molecular world, the methods pioneered by Perrin,
Svedberg, and others moreover opened the way to designing an effigy of
the molecular world on the colloidal scale
\cite{pusey:91,russel-saville-schowalter:91,frenkel:2000}. The
\emph{natural laboratory} of Gouy was successfully transformed into an
\emph{artificial microcosmos} \cite{haw:2002}, which obviously leads
us slightly off--topic, again. Svedberg's ultracentrifuge helped to
solve one of the big experimental challenges in colloid preparation:
to prepare samples that have sufficiently well--defined properties, in
particular a monodisperse size distribution to mimic as closely as
possible the sameness of atoms. This is hard work: starting with one
kilogram of gamboge, which is a gum resin named after Cambodia (from
where it was originally imported), Perrin ``obtained after several
months of daily operations a fraction containing several decigrams of
grains with a diameter of approximately three--quarters of a
thousandth of a millimetre'' \cite{perrin:nobel}.  But monodisperse
micron--sized particles were a crucial prerequisite for the progress
of colloid science.  With an enormous toolbox at hand, chemists are
now able to design almost any colloidal interactions one may wish for,
far beyond what can be found among naturally occuring molecules.

This gave experimentalists a formidable playground for testing highly
idealized theoretical pet--models for many--body problems, such as the
hard--sphere system and its various extensions. As anticipated by van
der Waals in the 19th century, much of the packing structure of
liquids can be understood from the hard--shpere like mutual repulsions
between their constituent molecules \cite{hansen-mcdonald:86}.  Hard
spheres also provide a beautifully simple example (paradigmatic for
more complicated soft condensed matter systems
\cite{chaikin-lubensky:95} including living cells \cite{herzfeld:96})
how macroscopic order emerges from microscopic chaos, namely the
stochastic Brownian motion of the individual spheres
\cite{chaikin:2000}. Numerical simulations
\cite{alder-wainwright:57,hoover-ree:68} and experiments
\cite{pusey-van_megen:86} have established that above a sharply
defined volume fraction $\phi=0.494$, Brownian fluctuations drive the
particles into a crystal, because they gain space to wobble around
when they arrange themselves on a (virtual) lattice\footnote{This
illustrates the paradox that global order may help to increase
individual freedom.}. Microgravity experiments \cite{zhu-etal:97}
suggest that this crystal is made out of randomly stacked hexagonally
close--packed planes. It becomes space filling at $\phi=0.545$ and
seems to develop a slight preference for face--centered cubic
organization if it is compressed towards closest packing
\cite{close-packing} at $\phi=0.74$. Colloidal crystallization has
gained additional interest from those trying to crystallize proteins
for structure analysis \cite{bergfors:99}. While, in this context,
colloids are currently still mostly playing their usual role as
visible model systems for globular proteins \cite{poon:97},
programmable protein expression might in the future well overcome
chemistry as the more powerful machinery for providing colloid physics
with the ultimate design precision.

If quickly quenched to volume fractions $\phi>0.58$, hard spheres will
provide yet another interesting and suprising feature closely related
to our central theme: by their Brownian motion they trap themselves in
an arrested non-equilibrium amorphous state that looks like a liquid
but feels like a solid, and which may substantially delay or even
impede crystallization: \emph{the colloidal glass}\footnote{To be fair
to the photon \cite{einstein:05p}, we note aside that not only
diffusing Brownian particles may come to a halt, but also diffusing
light. Recently, colloidal crystallization has gained some attention
as a promising method to produce photonic crytals \cite{john:91},
which may become an attractive technology in the future (currently all
the rage: hiding light under a bushel).}.  From mode--coupling theory
\cite{goetze:91} a very characteristic (self--generated) subdiffusive
behavior \cite{Fuchs94} during the critical slowing--down of the
Brownian motion near this transition was predicted. This was
beautifully confirmed (see e.~g.\ Refs.~\cite{Goetze92,Goetze99}) for
collective density fluctuations by light scattering
\cite{van_megen-underwood:93}, and for tagged particles by numerical
simulations \cite{kob:99}.  At least for hard--sphere like systems the
theory can be regarded as legitimate --- though, unfortunately, not
completely rigorous\footnote{The worst theory apart from all others
that have been tried from time to time \cite{cates:2003}.} ---
sucessor of Eq.~(\ref{eq:deq}) towards a more microscopic picture. It
starts from a microscopic expression for the normalized dynamic
structure factor $\phi_q(t)\equiv S_q(t)/S_q(0)$, cf.\
Eq.~(\ref{eq:sqt}), in terms of generalized fluctuating particle
currents $J$
\begin{equation}\label{eq:mct}
 \frac{\partial \phi_q(t)}{\partial t}= - D(q)q^2\phi_q(t) + \int_0^t
 d\tau \; \avg{J_q(t-\tau)J_q(0)}\phi_q(\tau) \;.
\end{equation}
For short times and $q\to0$, this reduces to the Fourier transform of
the diffusion equation. The kinetic coefficient $D(q)$ generalizes the
gradient diffusion coefficient $D=D(q\to0)$ to finite wave
vectors. For long times, it becomes effectively reduced due to the
term reminiscent of Eq.~(\ref{eq:gkr}), through which the system
remembers its past evolution (e.~g.\ its long--time tails). To make
practical use of Eq.~(\ref{eq:mct}), the microscopic memory kernel
containing the Brownian fluctuations $J$ has to be expressed in
terms of the hydrodynamic variables $\phi$, which involves the
uncontrolled mode--coupling approximation
\cite{Goetze92,bouchaud-etal:96,Kawasaki98}. Yet, after iterative solution,
Eq.~(\ref{eq:mct}) predicts in impressive detail how, upon increasing
either the attraction or the volume fraction, the simple Brownian
dynamics of a fluid suspension described by Eq.~(\ref{eq:deq}) gives
way to slow glassy dynamics, and eventually to two disticint
non--ergodicity transitions into two fundamentally different glassy
states \cite{pham-etal:2002,sciortino:2002}.

\section{Fluctuating manifolds: Brownian particles with soft 
  internal modes}
\label{sec:fluctuating_manifolds}

Just as it is commonplace that history repeats itself, re--inventing
the wheel is an everyday experience in science. So it could have been
expected that researchers studying red blood cells under the
miscrosope would be tempted to ascribe their flickering shape
undulations to animate causes, thereby repeating the initial
misinterpretation of Brownian motion as a sign of life by (most of)
its early discoverers (see Ref.~\cite{brochard-lennon:75} for a
discussion of the history). The story of Brownian motion was about to
repeat itself for Brownian particles with soft internal degrees of
freedom. Today, cell flickering is understood to be a physical
phenomenon \cite{brochard-lennon:75} that has an important biological
implication, tough: the reduction of unfavorable adhesion via the
so--called Helfrich repulsion \cite{helfrich-harbich:85} caused by the
thermal undulations of the cell membrane. Similarly, the thermal
conformational wiggling of the polymeric constituents of the
cytoskeleton has been pinpointed as the origin of some characeristic
viscoelastic
behavior~\cite{mackintosh-kaes-janmey:95,hinner-etal:98,gardel-etal:2003}.

\subsection*{\bf Brownian undulations \dots}

For fluctuating manifolds, not only the center of mass position or
overall orientation but an (ideally) infinite number of internal modes
are excited into a persistent dance by thermal forces. Brownian
fluctuations tend to ``crumple'' low dimensional manifolds
\cite{membrane:winterschool}. Their free and driven\footnote{Opening
the Pandora box of non--equilibrium one enters the vast topic of
\emph{pinned} or \emph{driven} fluctuating manifolds
\cite{halpin-healy_zhang:95,kardar:98,fisher:98,panja:2004} that
comprises a wealth of complex non--equilibrium phenomena in many areas
of condensed matter physics, which we will not pursue any further,
here.}  conformational dynamics is the subject of a field of
investigation that one might summarize by the term \emph{stochastic
elastohydrodynamics}. It is largely motivated by the current interest
in soft and biological materials. Examples of fluctuating manifolds
are flexible linear and branched polymers, microemulsions, foams,
liquid crystals and numerous other soft surfaces, interfaces and
membranes~\cite{doi-edwards:86,degennes:79,%
membrane:winterschool,safran:2003,seifert:97,lipowsky:91,%
lipowsky_sackmann:95a}, but also vortex matter, line liquids, magnetic
domain walls and the like~\cite{nelson:book,blatter:review}.

For the most natural case of an elastically uniform manifold of $n=1$
or $n=2$ dimensions~\cite{frey_nelson:91}, the force needed to excite
internal modes will grow as a power law with the mode number $q$,
equilibrium amplitudes $a_q$ and relaxation times will decrease as
$\avg{a_q^2}\propto q^{-\beta}$ and $\tau_q\propto q^{-\gamma}$,
respectively. For $\gamma>\beta-n$, this is quickly translated into a
subdiffusive\footnote{If this inequality is violated, one does
\emph{not} obtain superdiffusion but the dynamics is dominated by the
lowest mode.}  mean--square displacement
\begin{equation}\label{eq:spd}
  \avg{\delta \ve r^2(t)} 
  = 2\sum_q\avg{\ve a^2_q}(1-e^{t/\tau_q})
 \propto t^{(\beta-n)/\gamma}  \; 
\end{equation}
by transforming the sum into an integral.  This is the appropriate
generalization of Eq.~(\ref{eq:msd}) to real--space measurements of a
small labelled patch on the manifold, which have indeed been very
successfully applied to various biomolecules
\cite{gittes-etal:93,caspi-etal:98,legoff-etal:2002}. Similarly as for
simple Brownian particles, Eq.~(\ref{eq:spd}) determines the decay of
the dynamic structure factor for long times ($t \to \infty$), when
only the self--correlations of patches of dimension $q^{-1}$ matter,
via Eq.~(\ref{eq:sqt}). In contrast, for times shorter than the
relaxation time of modes of wavelength $q^{-1}$, the fluctuating
interference of light scattered from neighboring elements of the
manifold \emph{within} such patches has not yet averaged out and
dominates the dynamic decorrelation of the scattered intensity. While
the internal modes thus affect the decay \emph{law} for long times,
$S_q(t\to\infty)\propto e^{-{\cal D} q^2t^\alpha}$, they merely
renormalize the effective diffusion coefficient for short times,
$S_q(t\to 0)\propto e^{- {\cal D}_q t}$. Both limits encode
information about the material properties of the manifold and the
solvent in different combinations. (Which regime dominates the dynamic
structure factor is determined by the elasticity of the manifold. For
flexible manifolds the simple exponential behavior dominates, and for
stiff ones the stretched exponential tail.) Under favorable conditions
this can be exploited to probe these parameters
\cite{hohenadl-etal:99} in analogy to the simple example of
particle--sizing discussed above.  Fluctuating manifolds thus add a
new facet to the story of Brownian motion: subdiffusive behavior may
not only arise due to complicated surroundings but equally well from
(quite simple!)  internal elastic degrees of freedom of the tracer. In
particular in presence of external forces this opens up the
possibility of rich non-equilibium Brownian dynamics even for such
seemingly trivial objects as a thread in a Newtonian solvent
\cite{hallatschek-frey-kroy:2004}.  If the long--range hydrodynamic
self--interactions, mentioned above in connection with the long time
tails, need to be included, they lead to considerable complications,
as e.~g.\ for flexible polymers \cite{doi-edwards:86} and for membranes
\cite{wiese:2000}, even in equilibrium.

\subsection*{\bf \dots in complex fluids}

Increasing the complexity by one more step, one can combine both
extensions of simple Brownian motion, internal elastic modes and a
complex solvent. Consider a long (tagged) polymer embedded into a
dense melt or solution of other (identical) polymers. The question how
the polymer diffuses through the medium is closely (but by no means
trivially) related to the question how the whole medium is able to
flow. Both are addressed by \emph{reptation theory}
\cite{degennes:79,doi-edwards:86,mcleish:2002}, the prevailing
phenomenological idea how this involved many--body problem can be
disentangled and reduced to a description based on a single polymer
laterally confined to an \emph{effective tube}. The concept is popular
for its intuitive elegance and its ability to rationalize data for the
elastic modulus and the viscosity. One imagines the polymer as a
random walk caught in a cage shaped like a curly tube of mean--square
end--to--end distance $\avg{R^2}$ proportional to its length
$L$. Taking into account that it has to diffuse back and forth all the
way along the tube to completely disengage from it, and that the
diffusion coefficient $D_\|$ for diffusion along its own contour is
inversely proportional to $L$, one deduces the disengagement time
\begin{equation}\label{eq:reptation}
 \tau_r\propto L^2/D_\|\propto L^3\;.
\end{equation}
For shorter times, due to multiple inter--polymer collisions
\cite{schaefer:99}, semidilute solutions or melts respond elastically
to shear, like a rubber with an elastic modulus of $k_BT$ per
statistically independent segment of the tube.  These segments (of
length $\avg{R^2}/L$) act as its effective unit elements. At times
longer than $\tau_r$, entanglements can relax, and the whole medium
can flow like a liquid. The large time $\tau_r$ therefore also fixes
the ratio of the medium viscosity at long times to the shear modulus
at shorter times, which implies that polymer melts and solutions are
soft but tenacious, or ``gloop''.  Support for the tube, intuitive
starting point of the reptation model, comes from direct visual
observation of the Brownian motion of large biopolymers
\cite{kaes-strey-sackmann:94,smith-perkins-chu:95}.  Towards a deeper
understanding of how viscoelasticity emerges from the underlying basic
laws of Brownian motion, microscopic approaches (theory
\cite{chong-fuchs:2002} and simulations
\cite{everaers-etal:2004,wittmer-etal:2004}) are currently pursued.

\begin{figure}[htb]
\begin{center}
  \includegraphics[width=0.5\textwidth]{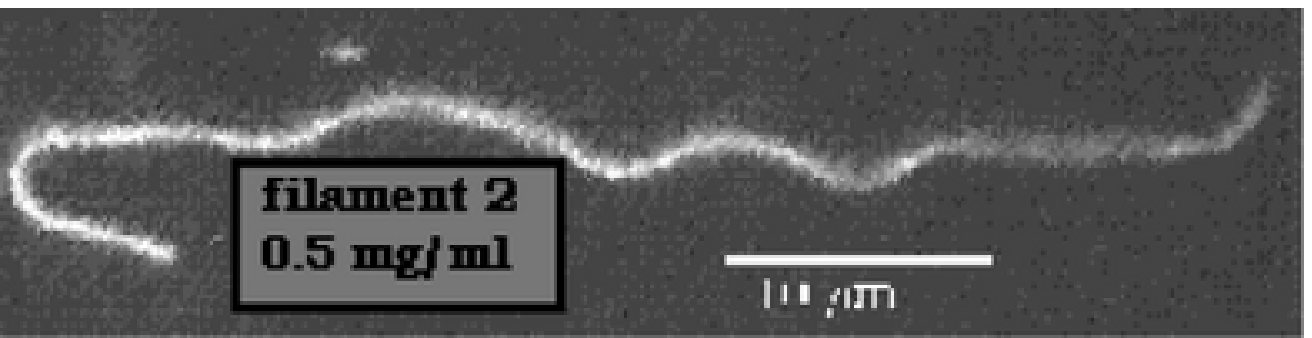}
  \hfill \includegraphics[width=0.4\textwidth]{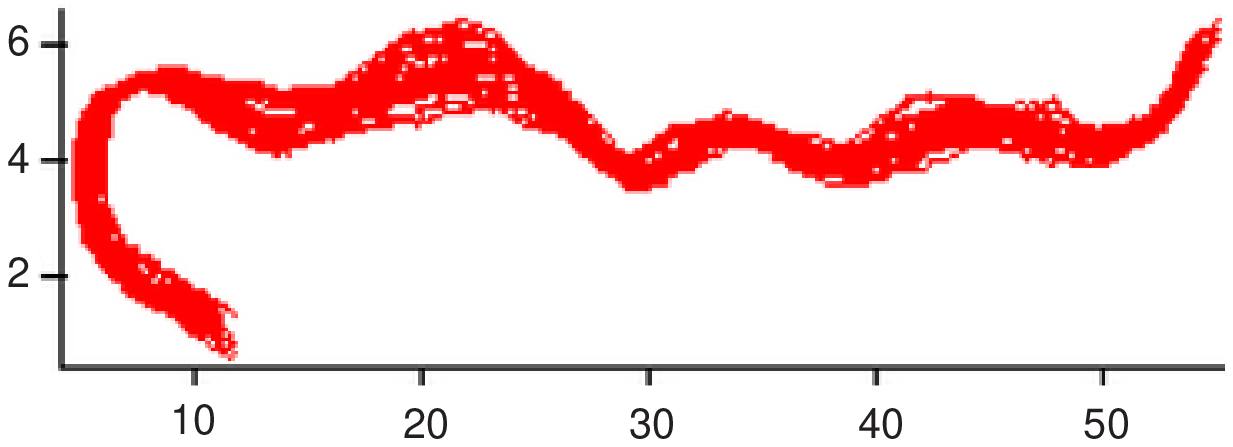}
\end{center}
\caption{Direct visualization of the tubes around actin filaments 
  in F-actin solutions.  Courtesy of J.
  K{\"a}s~\cite{kaes-strey-sackmann:94}.}
\label{fig:actin_reptation}
\end{figure}

Over the last decades, membranes and polymers have become the major
companions of the simple Brownian particle as paradigms of \emph{soft
condensed matter}, which is what we ourselves and many of our natural
and technological aides of everyday life (foods, cosmetics,
detergents, plastics \dots) are made of. Rather than continuing this
list ad infinitum, we close this section by pointing out an ironic
twist in the recent history of Brownian motion: with the physics of
fluctuating polymers and membranes reasonably well understood,
physicists have at last turned their attention back to the possible
animate sources of Brownian fluctuations. The idea is to study the
additional effect of ``active noise'', such as the undulations induced
by molecular pumps, molecular motors, and switching channels embedded
into a membrane~\cite{ramaswamy-toner-prost:2000,manneville_etal:99},
a theme that can be iterated in many directions; e.~g.\ towards
so--called active gels
\cite{nedelec-etal:97,humphrey_etal:02,uhde_etal:04,kruse-etal:2004,},
solutions of biopolymers mixed with molecular motors, in which rich
structure formation (asters and spirals, etc.) has been observed {\em
in vivo} and {\em in vitro}. Interestingly, the ``noise'' produced by
the biological activity was predicted to lead to a thousandfold
increase of the amplitudes of the long--time tails
\cite{hatwalne-etal:2004}.

\section{Rectified Brownian motion: from ``life force'' to ``living
  force'' and back} 
\label{sec:motors}

It is an amusing semantic detail from the history of science that
\emph{kinetic energy} was originally termed \emph{living force}
(``lebendige Kraft'') by Helmholtz and his contemporaries, adopting
Aristoteles' ``vis viva''.  When Brown first observed the irregular
motion of pollen granules, he was convinced that he had found animate
objects driven by some kind of ``life force'' outside the realm of
physics and chemistry, a vitalistic view still common at that time. He
derived from Leeuwenhoek's terminology of ``animalcules'' (small
animals) the name ``molecules'' for the rambling particles under his
microscope. After their animal aura had been stripped off by the
subsequent investigations, the term could be reused for those
particles that carried the thermal {\em living force} then recognized
as the mundane physical explanation of the hitherto mysterious {\em
  life force}.

This section will deal with the origin of motion in living systems and
will bring us back once again to the ``living force'' as the ``life
force'' responsible for directed motion on a cellular scale. The
origin of motion of living organisms was already debated in the 3rd
century B.C., when the anatomist and physician Erasistratos of Ceos
associated muscle motion with the "spiritus animalis". He imagined it
as some kind of fluid or gas, which he termed ``pneuma", that flows
through hollow nerves as pipelines and makes the muscles swell and
shorten.  This pneumatic idea survived in the scientific community for
a very long time. It was only with the invention of the microscope by
van Leeuwenhoek (1632-1723) that Swammerdam (1637-1680) was able to
show that muscles contract at constant volume which invalidated
pneumatic theories.  Helmholtz in his famous paper about the law of
energy conservation~\cite{helmholtz:lebendige_kraft} may have been the
first to emphasize that the mechanical energy produced by living
organisms is just transduced chemical energy~\footnote{``Es bleiben
  uns von den bekannten Naturprozessen noch die organischen Wesen
  {\"u}brig. ...  [Tiere] verbrauchen also eine gewisse Quantit{\"a}t
  chemischer Spannungskr{\"a}fte, und erzeugen daf{\"u}r W{\"a}rme und
  mechanische Kr{\"a}fte.'' At that time the term ``Spannungskraft''
  and ``lebendige Kraft'' stood for what we nowadays call potential
  and kinetic energy, respectively.}. These ideas paved the way for a
physical understanding of muscle contraction. A discussion of the
early work on theoretical models, which already contains some of the
ideas of more recent work discussed below, can be found in reviews of
A.F.  Huxley~\cite{huxley:review} and Hill~\cite{hill:review}.

Modern experimental techniques~\cite{mehta99a} have lifted the veils
and allowed us to look at the causes of biological motion on a
molecular scale. This resulted in the following
picture~\cite{howard:book}. Biological motion is caused by (nanometer
sized) motor proteins, a highly specialized class of enzymes which are
able to transduce the energy excess in the chemical hydrolysis
reaction of ATP (adenosine--triphosphate) into mechanical work. An
important subclass are cytoskeletal motors, which by sequence
similarity are classified into three families (myosins, kinesins and
dyneins). These motors ``walk'' along one-dimensional molecular tracks
consisting of quite stiff protein fibres (F-actin and microtubules).
They are involved in many biological processes essential for living
organisms, such as mitosis and meiosis, as well as muscle contraction
and intracellular transport.

\subsection*{\bf How individual motors work}
Looking at these systems with the eye of a theoretician, molecular
motors reduce to microscopic objects (and as such they are subject to
Brownian motion) moving uni-directionally along one-dimensional
periodic substrates. The obvious question is: ``What are the
mechanisms which rectify Brownian motion and can explain such a
uni-directional motion?'' The answer is that there are actually a
variety of mechanisms, different aspects of which are discussed in a
series of recent
reviews~\cite{haenggi_bartussek:96,astumian:review,juelicher_etal:review,%
  reimann:review,parmeggiani_schmidt:review} (see also the article by
P.  H{\"a}nggi {\it et al.} in this volume~\cite{haenggi_atal:05}).

\begin{figure}[htb]
\begin{center}
\includegraphics[width=0.4\textwidth] {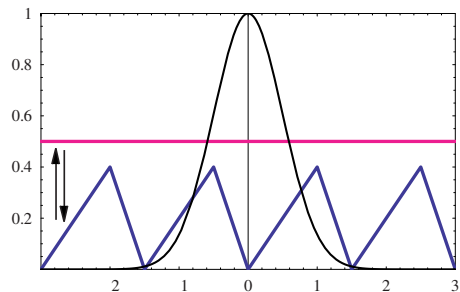}
\end{center}
\caption{Illustration of a ``flashing ratchet''. Upon switching a sawtooth-like
  potential on and off, a Brownian particle will move
  uni-directionally along a one-dimensional molecular track.}
\label{fig:brownian_ratchet}
\end{figure}

The essential idea can be summarized with the following rather
elementary but illustrative example, known as the ``flashing''
ratchet~\cite{bug_berne:87,ajdari92,astumian94}; see
Fig.~\ref{fig:brownian_ratchet}. Consider a Brownian particle subject
to viscous damping which is moving along a track with a periodic,
asymmetric potential $V(x)$ (with a sawtooth-like shape).  Despite the
spatial asymmetry of the potential, no preferential direction of
motion is possible if in an isothermal environment only equilibrium
fluctuations act on the particle. This argument goes back to
Smoluchowski~\cite{smoluchowski:1912} (see also the Feynman lectures
\cite{feynman_lectures:I}) who showed that otherwise one would be able
to construct a perpetuum mobile and violate the second law of
thermodynamics.

Thus, for generating uni-directional motion we obviously need an
additional ingredient which drives the system out of equilibrium.  In
a ``flashing'' ratchet this is achieved by cyclically turning the
potential on and off. The principle is easily explained.  Say that the
potential barriers are rather high compared to any thermal energy
scale. Then, with the potential switched on, the Brownian particle is
well localized in one of the potential minima. Upon turning the
potential off, the Brownian particle starts performing a free
one-dimensional random walk, such that after some time $\tau_{\rm
  off}$ the probability distribution $P(x)$ for finding it at position
$x$ is given by a Gaussian
\begin{equation}
P(x) = \frac{1}{\sqrt{4\pi D \tau_{\rm off}}}
       \exp\biggl[ - \frac{x^2}{4 D \tau_{\rm off}} \biggr] \, ,
\end{equation}
where $D$ is the diffusion coefficient. When the potential is turned
on again, the asymmetry of the potential acts as a rectifyer on the
probability distribution function and causes net transport in one
direction; see Fig.~\ref{fig:brownian_ratchet}. In this sense Brownian
motion acts as a ``life force'' (``Lebenskraft'') for molecular scale
engines. Note, however, that the
energy to drive the transport does not come from the thermal noise but
is provided when the potential is switched on and off.

Recently there have been several experimental ramifications of the
``flashing ratchet'' using colloidal
particles~\cite{rousselet_etal:94,faucheux_etal:95}. With the advances
in nanosciences and microfluidics there are by now many cleverly
designed systems, classical as well as quantum, which utilize Brownian
motion for generating transport, sorting particles and performing
other tasks~\cite{sarikay_etal:03}.

The ``flashing ratchet'' is, of course, not a realistic model for any
kind of actual biological motor. It oversimplifies a variety of
features of real biological engines. First, molecular motors are not
structureless particles but proteins with a quite complex architecture
and hence a large number of internal degrees of freedom. For example,
kinesin is a dimer consisting of two globular head domains joined
together by alpha-helical coiled-coil. Second, the interactions of the
motor protein with its fuel ATP and the molecular track are complex
molecular processes, a full analysis of which would require an
atomistic description including all the different types of forces
between the chemical agents in a watery environment. To go beyond
generic models making the proof of principle one has to design more
elaborate models which account for the most important structural and
chemical elements. What ``most important'' means varies from system to
system. It depends on specific features of the particular engine, e.~g.\
the type of coupling between the biochemical cycle and the
conformational states of the protein~\footnote{It is impossible to
  list the multitude of theoretical and experimental work in this
  area. The reader may consult some of the recent
  reviews~\cite{astumian:review,juelicher_etal:review,%
    reimann:review,parmeggiani_schmidt:review} and
  books~\cite{howard:book} for a more elaborate discussion and
  additional references.}. Depending on the level of coarse-graining
we may describe the forces acting on the degrees of freedom as
systematic or stochastic.  Systematic forces include viscous drag and
``mechanical'' forces acting on structural elements of the protein
which are sometimes visualized as parts of macroscopic engines
(springs, levers, joints, etc.). Stochastic forces may be classified
as Brownian conformational fluctuations and stochastic chemical
reactions. Both, of course, are just variations of the same theme.
Their common cause is the chaotic dance at a molecular scale. The
``flashing ratchet'' and the ``Possion stepper'' emphasize these two
sides of the medal. In the flashing ratchet all stochasticity results
from Brownian conformational fluctuations in a potential landscape. In
contrast, for the ``Poisson stepper'' (a textbook example for a
discrete stochastic process), which advances step by step at constant
rate, stochasticity results from the distribution of waiting times
between the steps~\cite{svoboda94b}.

\subsection*{\bf Traffic jams in the cell}

\begin{figure}[htb]
\begin{minipage}[t]{.45\textwidth}
\includegraphics[width=\textwidth]{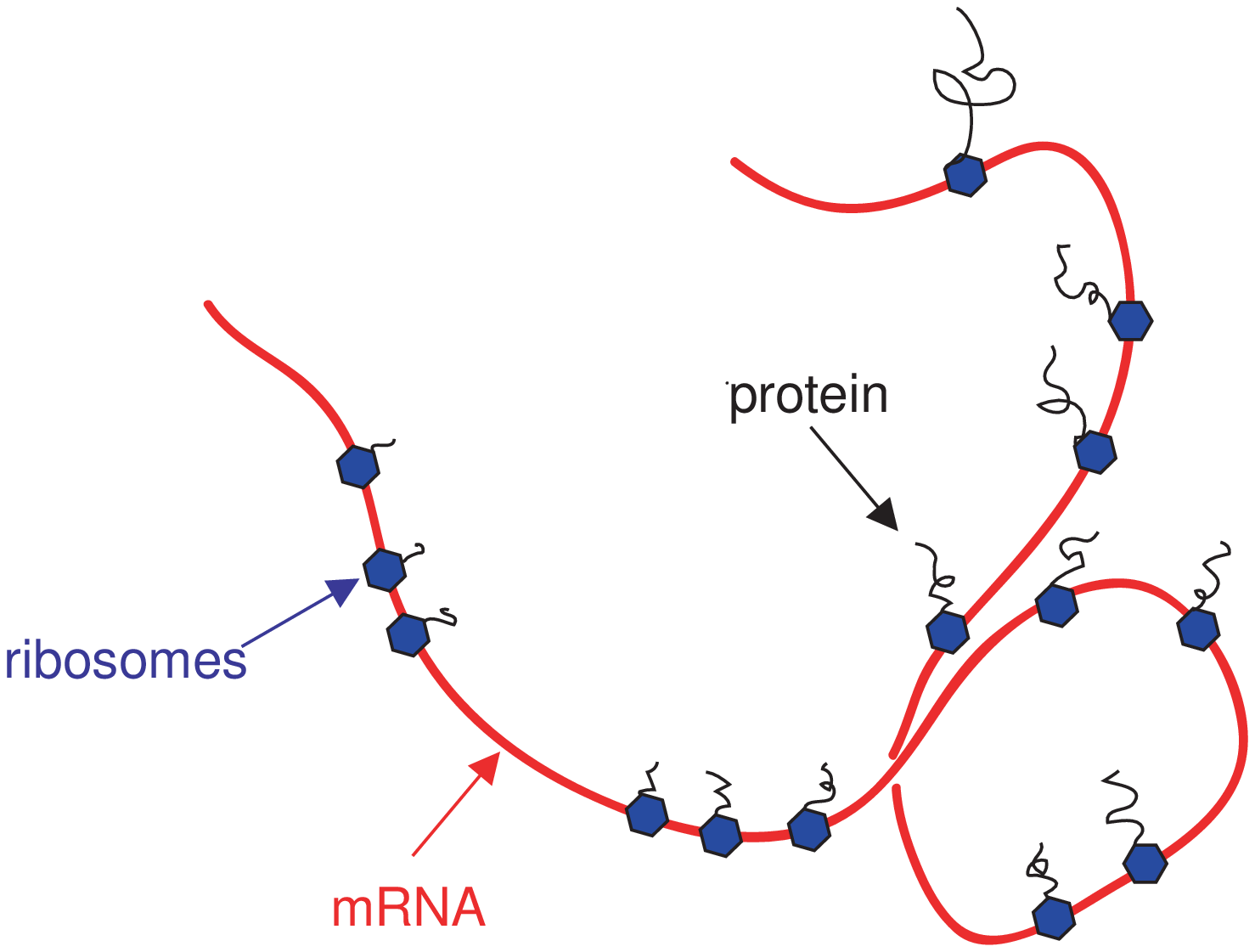}
\caption{Illustration of protein synthesis: ribosomes move
  from codon to codon along a mRNA template, reading off genetic
  information and thereby generating proteins step by
  step~\cite{alberts}}
\label{fig:mRNA}
\end{minipage}
\hfil
\begin{minipage}[t]{.45\textwidth}
\includegraphics[width=\textwidth]{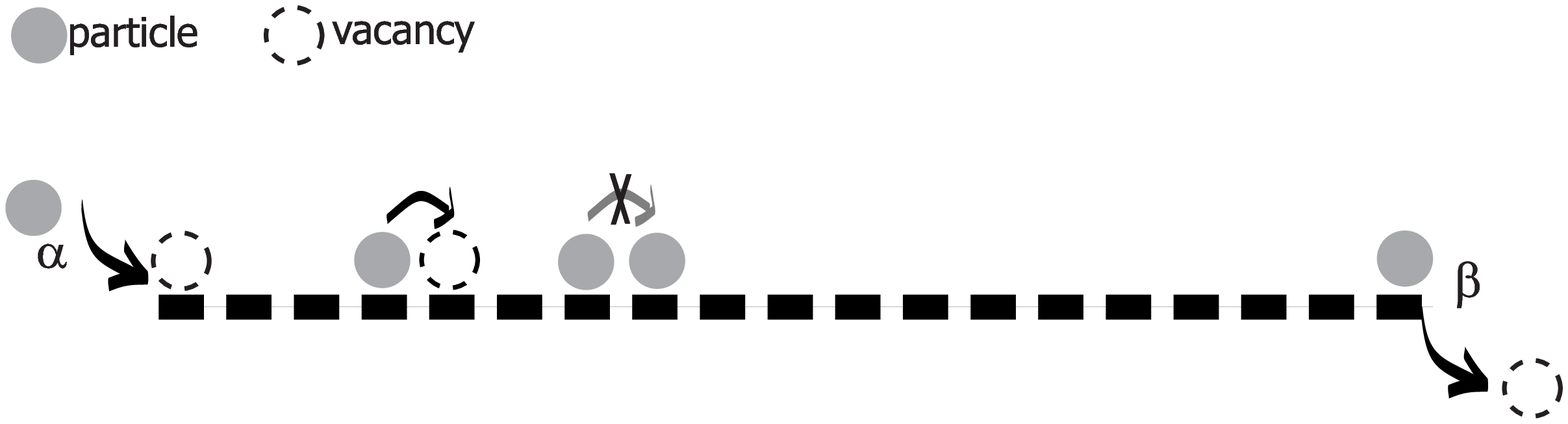}
\caption{Illustration of the Totally Asymmetric Simple Exclusion
  Process with open boundaries. The entrance and exit rates at the
  left and right end of the one-dimensional lattice are given by
  $\alpha$ and $\beta$, respectively.}
\label{fig:tasep}
\end{minipage}
\end{figure}

There are several biological processes where a concerted action of
molecular motors is of importance. A particulary prominent example is
protein sythesis by mRNA translation in eucaryotes, which involves
unidirectional motion of ribosome complexes along mRNA strands.
Theoretical investigations go back to the pioneering work by MacDonald
et al.~\cite{macdonald_gibbs_pipkin:68}, who designed a driven lattice
gas model which by now is known as the totally asymmetric exclusion
process (TASEP). In this model a single species of particles is
hopping unidirectionally and with a uniform rate along a
one-dimensional lattice.  The only interaction between the particles
is hard-core repulsion, which prevents more than one particle from
occupying the same site on the lattice; see Fig.~\ref{fig:tasep}.
Originally intended as a model for an important biological process,
this model has by now become one of the paradigms of non-equilibrium
physics~(for a review see
Refs.~\cite{spohn:book,derrida_evans:review,mukamel:review,schuetz:review}).

Much of what we know about the dynamics and the steady state of the
TASEP and driven lattice gases~\cite{schmittmann_zia:95} in general is
the result of theoretical work using a broad variety of tools and
exploiting the intimate connections between classical and quantum
physics, which we eluded to in section \ref{sec:methods}.  It has been found
that the nature of the non-equilibrium steady state of the TASEP
depends sensitively on the boundary conditions. For periodic boundary
conditions the system reaches a steady state of constant density.
Interestingly, density fluctuations are found to spread faster than
diffusively~\cite{beijeren_kutner_spohn:85}. This can be understood by
an exact mapping~\cite{meakin_etal:86} to a growing interface model,
whose dynamics in the continuum limit is described in terms of the
Kardar-Parisi-Zhang equation~\cite{kardar_parisi_zhang:86} and its
cousin the noisy Burgers equation~\cite{forster-nelson-stephen:77}.
In contrast to such ring systems, open systems with particle
reservoirs at the ends exhibit phase transitions upon varying the
boundary conditions~\cite{krug:91}.  This is genuinely different from
thermal equilibrium systems where boundary effects usually do not
affect the bulk behavior and become negligible if the system is large
enough. In addition, general theorems do not even allow equilibrium
phase transitions in one-dimensional systems at finite temperatures
(if the interactions are not too
long-range)~\cite{landau_lifshitz_stat_mech:book}.  Yet another
difference between equilibrium and non-equilibrium processes can be
clearly seen on the level of its dynamics. If transition rates between
microscopic configurations are obeying detailed balance the system is
guaranteed to evolve into thermal equilibrium. Systems lacking
detailed balance may still reach a steady state, but at present there
are no universal concepts like the Boltzmann-Gibbs ensemble theory for
characterizing such non-equilibrium steady states.  In most instances
one has to resort on solving nothing less than its full dynamics. It
is only recently, that exact (non-local) free energy functionals for
driven diffusive systems have been
derived~\cite{derrida_lebowitz_speer:01,derrida_lebowitz_speer:02}.

Recently, driven lattice gas models have found their way back into
biological physics. MacDonald's original ideas have been generalized
taking into account the finite extension of
ribosomes~\cite{shaw_zia_lee:03,chou:03}. Also some aspects of
intracellular transport, where the molecular motor kinesin moves along
microtubules, show close resemblance with the TASEP.  As an additional
feature one has to account for the fact that microtubules are embedded
in a cellular environment with a reservoir of motors in solution. This
allows for the motors to attach from the solution to the molecular
track and to detach from it to join the
reservoir~\cite{parmeggiani_franosch_frey:03,klumpp_lipowsky:03}.
Typically kinetic rates are such that these motors walk a significant
fraction along the molecular track before detaching. It came as quite
a surprise that the competition between the non-conservative
on/off-dynamics and the driven transport along the molecular track
leads to a separation between high and low density
phases~\cite{parmeggiani_franosch_frey:03}.

\section{Conclusions and Outlook}

After 100 years many of the central questions surrounding the
ubiquitious phenomenon of Brownian motion are still unresolved so that
it seems still too early for closing the books and drawing final
conclusions. As soon as we leave the caved paths of equilibrium
statistical mechanics traced out a century ago, intriguing open
problems and promising perspectives abound.  Still, we cannot derive
the basic assumptions of the founding fathers from the underlying
microscopic dynamics, and still we have not managed to generalize
reliably the powerful theoretical framework we inherited from them to
processes far from equilibrium.  We are only starting to appreciate to
what extent the processes of life are shaken by fluctuations, just
like the pollen grains studied by Brown almost 200 years ago. Life
undeniably shows a strong preference for the borderline between
perfect order and the erratic dance called Brownian motion, which
will, for the foreseeable future, remain a paradigm of the natural
sciences in a noisy world.

\begin{acknowledgement}
  We are indepted to Panayotis Benetatos, Matthias Fuchs, and Josef
  K\"as for discussiolns and valuable suggestions. We thank the
  authors of Refs.~\cite{maier_raedler:99,kaes-strey-sackmann:94} for
  permitting the reproduction of Figs.~\ref{fig:DNA} and
  \ref{fig:actin_reptation}.
\end{acknowledgement}


\begin{thebibliography}{100}

\bibitem{feynman_lectures:I}
R.~P. Feynman, R.~B. Leighton, and M.~L. Sands, { The Feynman Lectures on
  Physics} (Addison-Wesley, Reading, 1963), Vol.~I.

\bibitem{einstein:05}
A. Einstein, Ann. Phys. (Leipzig) {\bf 17},  549  (1905).

\bibitem{brown:28}
R. Brown, Edin. New Phil. J. {\bf 5},  358  (1828).

\bibitem{brongniart:27}
A. Brongniart, Ann. Sci. Naturelles {\bf 12},  41  (1827).

\bibitem{perrin:nobel}
J.~B. Perrin, Discontinuous Structure of Matter, Nobel Lecture, 1926.

\bibitem{bachelier:00}
L. Bachelier, Annales Scientifiques de {l'\'E}cole {N}ormale {S}up{\'e}rieure
  {\bf 17},  21  (1900).

\bibitem{dunbar:2000}
N. Dunbar, { Inventing Money: The Story of Long-Term Capital Management and
  the Legends Behind It} (Wiley, Chichester, 2000).

\bibitem{smoluchowski:06}
M. von Smoluchowski, Ann. Phys. (Leipzig)  756  (1906).

\bibitem{langevin:08}
P. Langevin, Comptes. Rendues {\bf 146},  530  (1908).

\bibitem{haw:2005}
M. Haw, Physics World  (2005), to appear.

\bibitem{coleman:2003}
P. Coleman, Ann.~Henri Poincar{\'e} {\bf 4},  1  (2003).

\bibitem{antunes-bettencourt-zurek:99}
N.~D. Antunes, L.~M.~A. Bettencourt, and W.~H. Zurek, Phys.~Rev.~Lett. {\bf
  82},  2824  (1999).

\bibitem{volovik:2001}
G.~E. Volovik, Phys.~Rep. {\bf 351},  195  (2001).

\bibitem{anderson:72}
P.~W. Anderson, Science {\bf 177},  393  (1972).

\bibitem{chaikin-lubensky:95}
P.~M. Chaikin and T.~C. Lubensky, { Principles of condensed matter physics}
  (Cambridge University Press, Cambridge, 1995).

\bibitem{cates-evans:2000}
{ Soft and Fragile Matter: Nonequilibrium Dynamics, Metastability and Flow},
  edited by M.~E. Cates and M.~R. Evans (SUSSP {$\&$} IOP, Bristol, 2000).

\bibitem{denk-watt:89}
W. Denk and W.~W. Webb, Phys.~Rev.~Lett. {\bf 63},  207  (1989).

\bibitem{saulson:90}
P.~R. Saulson, Phys.~Rev.~D {\bf 42},  2437  (1990).

\bibitem{kerner:2004}
B.~S. Kerner, { The Physics of Traffic. Empirical Freeway Pattern Features,
  Engineering Applications, and Theory}, { Springer Series Understanding
  Complex Systems} (Springer, Berlin, 2004).

\bibitem{schweitzer:2003}
F. Schweitzer, { Brownian Agents and Active Particles. Collective Dynamics
  in the Natural and Social Sciences}, { Springer Series in Synergetics}
  (Springer, Berlin, 2003).

\bibitem{challet-etal:2001}
D. Challet, A. Chessa, M. Marsili, and Y.-C. Zhang, Quantitative Finance {\bf
  1},  168  (2001).

\bibitem{mantegna-stanley:2000}
R.~N. Mantegna and H.~E. Stanley, { An Introduction to Econophysics:
  Correlations and Complexity in Finance} (Cambridge University Press,
  Cambridge, 2000).

\bibitem{bouchaud-potters:2004}
J.-P. Bouchaud and M. Potters, { Theory of Financial Risk and Derivative
  Pricing: From Statistical Physics to Risk Management}, 2nd ed. (Cambridge
  University Press, Cambridge, 2004).

\bibitem{neumann_morgenstern:book}
J. von Neumann and O. Morgenstern, { Theory of Games and Economic Behavior}
  (Princeton University Press, Princeton, 1944).

\bibitem{haw:2002}
M. Haw, J.~Phys.:~Condens.~Matter {\bf 14},  7769  (2002).

\bibitem{chandrasekhar:43}
S. Chandrasekhar, Rev.~Mod.~Phys. {\bf 15},  1  (1943).

\bibitem{nelson:67}
E. Nelson, { Dynamical Theories of Brownian Motion} (Princeton Univerisity
  Press, Princeton, 1967), $2^{\rm nd}$ ed. at
  http://www.math.princeton.edu/~nelson/books.html.

\bibitem{gardiner:85}
C.~W. Gardiner, { Handbook of Stochastic Methods for Physics, Chemistry and
  the Natural Sciences}, Vol.~13 of { Springer Series in Synergetics}
  (Springer, Berlin, 1985).

\bibitem{mazo:2002}
R.~M. Mazo, { Brownian Motion. Fluctuations, Dynamics, and Applications}
  (Clarendon, New York, 2002).

\bibitem{mermin:2004}
N.~D. Mermin, Physics Today {\bf 57},  10  (2004).

\bibitem{gouy:88}
L. Gouy, J. Physique {\bf 7},  561  (1888).

\bibitem{wiener:63}
C. Wiener, Ann. Phys. {\bf 118},  79  (1863).

\bibitem{agutter-malone-wheatley:2000}
P.~S. Agutter, P.~C. Malone, and D.~N. Wheatley, J.~Hist.~Biol. {\bf 33},  71
  (2000).

\bibitem{fick:1855}
A. Fick, Ann. Phys. (Leipzig) {\bf 94},  59  (1855).

\bibitem{einstein:06}
A. Einstein, Ann. Phys. (Leipzig) {\bf 19},  371  (1906).

\bibitem{feynman-hibbs:65}
R.~P. Feynman and A.~R. Hibbs, { Quantum Mechanics and Path Integrals}
  (McGraw-Hill, New York, 1965).

\bibitem{einstein:07}
A. Einstein, Ann. Phys. (Leipzig) {\bf 22},  569  (1907).

\bibitem{einstein:10}
A. Einstein, Ann. Phys. (Leipzig) {\bf 33},  1275  (1910).

\bibitem{irons:2004}
F.~E. Irons, Am.~J.~Phys. {\bf 72},  1059  (2004).

\bibitem{fluctuation-phenomena:87}
{ Fluctuation Phenomena}, 2 ed., edited by E.~W. Montroll and J.~L. Lebowitz
  (Elsevier, Amsterdam, 1987).

\bibitem{kubo:86}
R. Kubo, Science {\bf 233},  330  (1986).

\bibitem{foelsing:97}
A. F{\"o}lsing, { Albert Einstein: a biography} (Penguin Books, London,
  1997).

\bibitem{johnson:28}
J.~B. Johnson, Phys.~Rev. {\bf 32},  97  (1928).

\bibitem{nyquist:28}
H. Nyquist, Phys.~Rev. {\bf 32},  110  (1928).

\bibitem{domb:96}
C. Domb, { The Critical Point: A historical introduction to the modern
  theory of critical phenomena} (Taylor {$\&$} Francis, London, 1996).

\bibitem{hohenberg-halperin:77}
P.~C. {Hohenberg} and B.~I. {Halperin}, Rev.~Mod.~Phys. {\bf 49},  435  (1977).

\bibitem{frey_schwabl:94}
E. Frey and F. Schwabl, Adv. in Phys. {\bf 53},  577  (1994).

\bibitem{wiegand:2004}
S. Wiegand, J.~Phys.:~Condens.~Matter {\bf 16},  R357  (2004).

\bibitem{kadanoff-martin:63}
L.~P. Kadanoff and P.~C. Martin, Annals of Physics {\bf 24},  419  (1963).

\bibitem{forster:75}
D. Forster, { Hydrodynamic Fluctuations, Broken Symmetry, and Correlation
  Functions} (Addison Wesley, Redwood City, 1975).

\bibitem{brenig:89}
W. Brenig, { Statistical Theory of Heat: Nonequilibrium Phenomena}
  (Springer, Heidelberg, 1989).

\bibitem{zwanzig:2001}
R. Zwanzig, { Nonequilibrium Statistical Mechanics} (Oxford University
  Press, Oxford, 2001).

\bibitem{dorfman:99}
J.~R. Dorfman, { An Introduction to Chaos in NonEquilibrium Statistical
  Mechanics} (Cambridge University Press, Cambridge, 1999).

\bibitem{gallavotti-bonetto-gentile:2004}
G. Gallavotti, F. Bonetto, and G. Gentile, { Aspects of Ergodic, Qualitative
  and Statistical Theory of Motion} (Springer, Berlin, 2004).

\bibitem{bunimovich-sinai:81}
L. Bunimovich and Y.~G. Sina{\u i}, Commun.~Math.~Phys. {\bf 78},  479  (1981).

\bibitem{gaspard:98}
P. Gaspard, { Chaos, Scattering, and Statistical Mechanics} (Cambridge
  University Press, Cambridge, 1998).

\bibitem{bonetto-lebowitz-rey_bellet:tbp}
F. Bonetto, J.~L. Lebowitz, and L. Rey-Bellet, math-ph/0002052, submitted to
  World Scientific (unpublished).

\bibitem{cecconi-etal:2005}
F. Cecconi, M. Cencini, M. Falcioni, and A. Vulpiani,  {\bf xx},  xx  (2005),
  to appear.

\bibitem{gaspard-etal:98}
P. Gaspard, M.~E. Briggs, M.~K. Francis, J.~V. Sengers, R.~W. Gammon, J.~R.
  Dorfman, and R.~V. Calabrese, Nature {\bf 394},  865  (1998).

\bibitem{schmittmann_zia:95}
B. Schmittmann and R. Zia,  in { Phase Transitions and Critical Phenomena},
  edited by C. Domb and J. Lebowitz (Academic Press, London, 1995), Vol.~17.

\bibitem{berthier-barrat:2002}
L. Berthier and J.-L. Barrat, Phys.~Rev.~Lett. {\bf 89},  095702  (2002).

\bibitem{cipelletti-etal:2000}
L. Cipelletti, S. Manley, R.~C. Ball, and D.~A. Weitz, Phys.~Rev.~Lett. {\bf
  84},  2275  (2000).

\bibitem{fabry-etal:2001}
B. Fabry, G.~N. Maksym, J.~P. Butler, M. Glogauer, D. Navajas, and J.~J.
  Fredberg, Phys.~Rev.~Lett. {\bf 87},  148102  (2001).

\bibitem{liphardt-etal:2002}
J. Liphardt, S. Dumont, S.~B. Smith, I. Tinoco~Jr., and C. Bustamente, Science
  {\bf 296},  1832  (2002).

\bibitem{taeuber:notes}
U. T{\"a}uber, Critical dynamics, http://www.phys.vt.edu (2003).

\bibitem{jarzynski:2004}
C. Jarzynski, Nonequilibrium work theorem for a system strongly coupled to a
  thermal environment, 2004, cond-mat/0407340.

\bibitem{les_houches:2003}
 in { Slow relaxations and nonequilibrium dynamics in condensed matter},
  Vol.~Session LXXVII (2002) of { Les Houches Summer Schools of Theoretical
  Physics}, edited by J.-L. Barrat, M. Feigelman, J. Kurchan, and J. Dalibard
  (EDP Sciences {$\&$} Springer, Berlin, 2003), pp.\ 367--521.

\bibitem{sollich-etal:97}
P. Sollich, F. Lequeux, P. H{\'e}braud, and M.~E. Cates, Phys.~Rev.~Lett. {\bf
  78},  2020  (1997).

\bibitem{young:97}
{ Spin Glasses and Random Fields}, edited by A.~P. Young
  (World Scientific, Singapore, 1997).

\bibitem{buchanan:2003}
M. Buchanan, Nature {\bf 425},  556  (2003).

\bibitem{iben-etal:89}
I.~E.~T. {Iben}, D. {Braunstein}, W. {Doster}, H. {Frauenfelder}, M.~K. {Hong},
  J.~B. {Johnson}, S. {Luck}, P. {Ormos}, A. {Schulte}, P.~J. {Steinbach},
  A.~H. {Xie}, and R.~D. {Young}, Physical Review Letters {\bf 62},  1916
  (1989).

\bibitem{bouchaud-georges:90}
J.-P. Bouchaud and A. Georges, Phys.~Rep. {\bf 195},  127  (1990).

\bibitem{bouchaud-etal:97}
J.-P. Bouchaud, L.~F. Cugliandolo, J. Kurchan, and M. Mezard,  in {
  Spin--glasses and random fields}, edited by P. Young (World Scientific,
  Singapore, 1997), pp.\ 161--223.

\bibitem{debenedetti-stillinger:2001}
P.~G. Debenedetti and F.~H. Stillinger, Nature {\bf 410},  259  (2001).

\bibitem{crisanti-ritort:2003}
A. Crisanti and F. Ritort, J.~Phys.~A. {\bf 36},  R181  (2003).

\bibitem{cugliandolo-kurchan:93}
L.~F. Cugliandolo and J. Kurchan, Phys.~Rev.~Lett. {\bf 71},  173  (1993).

\bibitem{szamel:2004}
G. Szamel, Phys.~Rev.~Lett. {\bf 93},  178301  (2004).

\bibitem{gross:2004}
D. Gross, Physics World  23  (2004).

\bibitem{gallavotti-cohen:95}
G. Gallavotti and E.~G.~D. Cohen, Phys.~Rev.~Lett. {\bf 74},  2694  (1995).

\bibitem{jarzynski:98}
C. Jarzynski, Phys.~Rev.~Lett. {\bf 78},  2690  (1998).

\bibitem{ruelle:2004}
D. Ruelle, Physics Today {\bf 57},  48  (2004).

\bibitem{evans-searles:2002}
D.~J. Evans and D.~J. Searles, Adv. Phys. {\bf 51},  1529   (2002).

\bibitem{gallavotti:2002}
G. Gallavotti, { Foundations of Fluid Dynamics} (Springer, Berlin, 2002).

\bibitem{maes-netocny:2003}
C. Maes and K. Netocn{\'y}, J. Stat. Phys. {\bf 110},  269  (2003).

\bibitem{crooks:2000}
G.~E. {Crooks}, Phys.~Rev.~E {\bf 61},  2361  (2000).

\bibitem{ritort:2003}
F. Ritort, Poincar{\'e} Seminar {\bf 2},  195  (2003).

\bibitem{wang-etal:2002}
G.~M. {Wang}, E.~M. {Sevick}, E. {Mittag}, D.~J. {Searles}, and D.~J. {Evans},
  Phys.~Rev.~Lett. {\bf 89},  050601  (2002).

\bibitem{van_zon-cohen:2003}
R. van Zon and E.~G.~D. Cohen, Phys.~Rev.~Lett. {\bf 91},  110601  (2003).

\bibitem{baldassarri-colaiori-castellano:2003}
A. Baldassarri, F. Colaiori, and C. Castellano, Phys.~Rev.~Lett. {\bf 90},
  060601  (2003).

\bibitem{ehrhardt-majumdar-bray:2004}
G. Ehrhardt, S. Majumdar, and A. Bray, Phys.~Rev.~E {\bf 69},  016106  (2004).

\bibitem{feller:68}
W. Feller, { An Introduction to Probability Theory and Its Applications}
  (Wiley, London, 1968), Vol.~1.

\bibitem{wiener:21}
N. Wiener, Proc.\ Natl.\ Acad.\ Sci.\ (USA) {\bf 7},  253, 294  (1921).

\bibitem{kleinert:2004}
H. Kleinert, { Path Integrals in Quantum Mechanics, Statistics, Polymer
  Physics, and Financial Markets}, 3 ed. (World Scientific, Singapore, 2004).

\bibitem{kac:49}
M. Kac, Trans. Am. Math. Soc. {\bf 65},  1  (1949).

\bibitem{degennes:72}
P. de~Gennes, Phys. Lett. A {\bf 38},  339  (1972).

\bibitem{doi-edwards:86}
M. Doi and S.~F. Edwards, { The Theory of Polymer Dynamics} (Clarendon
  Press, Oxford, 1986).

\bibitem{descloizeaux:90}
J. des Cloizeaux and G. Jannink, { Polymers in Solution} (Clarendon Press,
  Oxford, 1990).

\bibitem{membrane:winterschool}
{ Statstical Mechanics of Membranes and Surfaces}, Vol.~5 of { Jerusalem
  Winter School for Theoretical Physics}, edited by D. Nelson, T. Piran, and S.
  Weinberg (World Scientific, Singapore, 1989).

\bibitem{schaefer:99}
L. Sch{\"a}fer, { Excluded Volume Effects in Polymer Solutions as Explained
  by the Renormalization Group} (Springer, Berlin, 1999).

\bibitem{maier_raedler:99}
B. Maier and J. R{\"a}dler, Phys.~Rev.~Lett. {\bf 82},  1911  (1999).

\bibitem{frey-etal:97} E. Frey, K. Kroy, J. Wilhelm, and E. Sackmann,
  in {Statistical Mechanics of Semiflexible Polymers: Theory and
    Experiments}, edited by D. Beysens and G. Forgacs (EDP Sciences,
  Springer Verlag, Berlin 1997), pp. 103.

\bibitem{bustamante-bryant-smith:2003}
C. Bustamante, Z. Bryant, and S. Smith, Nature {\bf 421},  423  (2003).

\bibitem{peyrard:2004}
M. Peyrard, Nonlinearity {\bf 17},  R1  (2004).

\bibitem{nelson_david:2004}
D.~R. Nelson, Statistical Physics of Unzipping DNA, 2004, cond-mat/0309559.

\bibitem{taeuber:03}
U. T{\"a}uber, Adv. Sol. State Phys. {\bf 43},  659  (2003).

\bibitem{sasai_wolynes:03}
M. Sasai and P. Wolynes, Proc.\ Natl.\ Acad.\ Sci.\ (USA) {\bf 100},  2374
  (2003).

\bibitem{kimura:83}
M. Kimura, { The neutral theory of molecular evolution} (Cambridge
  University Press, Cambridge, 1983).

\bibitem{fisher:22}
R. Fisher, Proc. Roy. Soc. Edinburgh {\bf 42},  321  (1922).

\bibitem{griffiths_etal:book}
A. Griffiths, J. Miller, D. Suzuki, R. Lewontin, and W. Gelbart, { An
  Introduction to Genetic Analysis}, 7 ed. (W.H. Freeman, New York, 1999).

\bibitem{liggett:book}
T. Liggett, { Interacting Particle Systems} (Springer Verlag, New York,
  1985).

\bibitem{krapivsky_redner:96}
P. Krapivsky and S. Redner, Am. J. Phys. {\bf 64},  546  (1996).

\bibitem{Ben-Avraham_etal:90}
D. Ben-Avraham, D. Considine, P. Meakin, S. Redner, and H. Takayasu,
  J.~Phys.~A. {\bf 23},  4297  (1990).

\bibitem{kimura:55}
M. Kimura, Proc.\ Natl.\ Acad.\ Sci.\ (USA) {\bf 41},  144  (1955).

\bibitem{jones_smith:book}
P. Jones and P. Smith, { Stochastic Processes: An Introduction} (Oxford
  University Press, New York, 2001).

\bibitem{redner:book}
S. Redner, { A Guide to First-Passage Processes} (Cambridge University
  Press, Cambridge, 2001).

\bibitem{kampen:92}
N.~G. van Kampen, { Stochastic Processes in Physics and Chemistry} (Elsevier
  Science B. V., Amsterdam, 1992).

\bibitem{kimura_ohta:69}
M. Kimura and T. Ohta, Genetics {\bf 61},  763  (1969).

\bibitem{kazazian:04}
H. Kazazian, Science {\bf 303},  1626  (2004).

\bibitem{drossel:01}
B. Drossel, Adv. Phys. {\bf 50},  209  (2001).

\bibitem{baake_gabriel:00}
E. Baake and W. Gabriel, Ann. Rev. Comp. Phys. {\bf 7},  203  (2000).

\bibitem{kramers:40}
H.~A. Kramers, Physica {\bf 7},  284  (1940).

\bibitem{haenggi_talkner_borkovec:90}
P. H{\"a}nggi, P. Talkner, and M. Borkovec, Rev.~Mod.~Phys. {\bf 62},  251
  (1990).

\bibitem{delbrueck:40}
M. Delbr{\"u}ck, J.~Chem.~Phys. {\bf 8},  120  (1940).

\bibitem{toussaint_wilczek:83}
D. Toussaint and F. Wilczek, J. Chem. Phys. {\bf 78},  2642  (1983).

\bibitem{lee94}
B.~P. Lee, J.~Phys.~A. {\bf 27},  2633  (1994).

\bibitem{hinrichsen:00}
H. Hinrichsen, Adv. Phys. {\bf 49},  815  (2000).

\bibitem{onsager31}
L. Onsager, Phys.~Rev. {\bf 37},  405  (1931).

\bibitem{cardyreview}
J. Cardy,  in { Proceedings of mathematical beauty of physics}, Vol.~24 of
  { Advanced Series in Mathematical Physics}, edited by J.-B. Zuber (World
  Scientific, Singapore, 1997), p.\ 113, cond-mat/9607163.

\bibitem{guptasarma:95}
P. Guptasarma, Bioessays {\bf 17},  987  (1995).

\bibitem{spudich_koshland:76}
J.~L. Spudich and D.~E. Koshland, Nature {\bf 262},  467  (1976).

\bibitem{mcadams_arkin:97}
H. McAdams and A. Arkin, Proc.\ Natl.\ Acad.\ Sci.\ (USA) {\bf 94},  814
  (1997).

\bibitem{elowitz_etal:02}
M. Elowitz, A. Levine, E. Siggia, and P. Swain, Science {\bf 297},  1183
  (2002).

\bibitem{mcadams_arkin:99}
H. McAdams and A. Arkin, Trends Genet. {\bf 15},  65  (1999).

\bibitem{henri:08}
V. Henri, C. R. Acad. Sci., Paris {\bf 146},  1024  (1908).

\bibitem{amblard-etal:96}
F. Amblard, A.~C. Maggs, B. Yurke, A.~N. Pargellis, and S. Leibler,
  Phys.~Rev.~Lett. {\bf 77},  4470  (1996).

\bibitem{gittes-etal:97}
F. Gittes, B. Schnurr, P.~D. Olmsted, F.~C. MacKintosh, and C.~F. Schmidt,
  Phys.~Rev.~Lett. {\bf 79},  3286  (1997).

\bibitem{crocker-etal:2000}
J.~C. Crocker, M.~T. Valentine, E.~R. Weeks, T. Gisler, P.~D. Kaplan, A.~G.
  Yodh, and D.~A. Weitz, Phys.~Rev.~Lett. {\bf 85},  888  (2000).

\bibitem{forstner-etal:2003}
M.~B. Forstner, D.~S. Martin, A.~M. Navar, and J.~A. K{\"a}s, Langmuir {\bf
  19},  4876  (2003).

\bibitem{maret-wolf:87}
G. Maret and P.~E. Wolf, Z. Phys. B: Condens. Matter {\bf 65},  409  (1987).

\bibitem{weitz-pine:93}
D.~A. Weitz and D.~J. Pine,  in { Dynamic Light Scattering}, edited by W.
  Brown (Oxford University Press, New York, 1993), Chap.~16, pp.\ 652--720.

\bibitem{schurr:2000}
J.~M. Schurr, Biophys. J. {\bf 79},  1692  (2000).

\bibitem{rigler-elson:2001}
{ Fluorescence Correlation Spectroscopy. Theory and Applications}, Vol.~65
  of { Springer Series in Chemical Physics}, edited by R. Rigler and E.~S.
  Elson (Springer, Heidelberg, 2001).

\bibitem{bausch-moller-sackmann:99}
A.~R. Bausch, W. M{\"o}ller, and E. Sackmann, Biophys. J. {\bf 76},  573
  (1999).

\bibitem{heilbronn:22}
A. Heilbronn, Jahrb.~Wiss.~Bot. {\bf 61},  284  (1922).

\bibitem{shlesinger-zaslavsky-klafter:93}
M.~F. Shlesinger, G.~M. Zaslavsky, and J. Klafter, Nature {\bf 363},  31
  (1993).

\bibitem{mandelbrot:83}
B. Mandelbrot, { The Fractal Geometry of Nature} (Freeman, San Francisco,
  1983).

\bibitem{tsallis-etal:95}
C. {Tsallis}, S.~V.~F. {Levy}, A.~M.~C. {Souza}, and R. {Maynard}, Physical
  Review Letters {\bf 75},  3589  (1995).

\bibitem{metzler-klafter:2000}
R. Metzler and J. Klafter, Phys.~Rep. {\bf 339},  1  (2000).

\bibitem{weiss-etal:2004}
M. Weiss, M. Elsner, F. Kartberg, and T. Nilsson, Biophys. J. {\bf 87},  3518
  (2004).

\bibitem{tolic-noerrelykke:2004}
I.~M. Toli{\'c}-N{\o}rrelykke, E.-L. Munteanu, G. Thon, L. Oddershede, and K.
  Berg-S{\o}rensen, Phys.~Rev.~Lett. {\bf 93},  078102  (2004).

\bibitem{ott_etal:90}
A. Ott, J. Bouchaud, D. Langevin, and W. Urbach, Phys.~Rev.~Lett. {\bf 65},
  2201  (1990).

\bibitem{wong-etal:2004}
I.~Y. Wong, M.~L. Gardel, D.~R. Reichman, E.~R. Weeks, M.~T. Valentine, A.~R.
  Bausch, and D.~A. Weitz, Phys.~Rev.~Lett. {\bf 92},  178101  (2004).

\bibitem{ben_naim-redner-ben_avraham:92}
E. Ben-Naim, S. Redner, and D. {b}en Avraham, Phys.~Rev.~A {\bf 45},  7207
  (1992).

\bibitem{benavraham-havlin:2000}
D. ben Avraham and S. Havlin, { Diffusion and Reaction in Fractals}
  (Cambridge University Press, Cambridge, 2000).

\bibitem{stauffer-aharony:94}
D. Stauffer and A. Aharony, { Introduction to Percolation Theory}, 2 ed.
  (Taylor and Francis, London, 1994).

\bibitem{tzanakis-grecos:99}
C. Tzanakis and A. Grecos, Transport Theor. Stat. Phys. {\bf 28},  325  (1999).

\bibitem{mclennan:88} J.~A. McLennan, { Introduction to
    Non-Equilibrium Statistical Mechanics} (Prentice-Hall, New York,
  1988).

\bibitem{dorfman-cohen:65}
J.~R. Dorfman and E.~G.~D. Cohen, Phys.~Lett. {\bf 16},  124  (1965).

\bibitem{alder-wainwright:67}
B.~J. Alder and T.~E. Wainwright, Phys.~Rev.~Lett. {\bf 18},  988  (1967).

\bibitem{ernst-hauge-van_leeuwen:70}
M.~H. Ernst, E.~H. Hauge, and J.~M.~J. van Leeuwen, Phys.~Rev.~Lett. {\bf 25},
  1254  (1970).

\bibitem{keyes-masters:85}
T. Keyes and A.~J. Masters, Adv. Chem. Phys. {\bf 58},  1  (1985).

\bibitem{landau-lifshitz:fm}
L.~D. Landau and E.~M. Lifshitz, { Fluid Mechanics}, Vol.~6 of { Course
  of Theoretical Physics} (Pergamon Press, London, 1963).

\bibitem{evans-morriss:90}
D.~J. Evans and G.~P. Morriss, { Statistical Mechanics of NonEquilibrium
  Liquids} (Academic Press, London, 1990).

\bibitem{fuchs-kroy:2002}
M. Fuchs and K. Kroy, J.~Phys.:~Condens.~Matter {\bf 14},  9223  (2002).

\bibitem{wolynes:76}
P.~G. Wolynes, Phys.~Rev.~A {\bf 13},  1235  (1976).

\bibitem{bocquet-barrat:94}
L. Bocquet and J.-L. Barrat, Phys.~Rev.~E {\bf 49},  3079  (1994).

\bibitem{hagen-etal:97}
M.~H.~J. {Hagen}, I. {Pagonabarraga}, C.~P. {Lowe}, and D. {Frenkel}, Physical
  Review Letters {\bf 78},  3785  (1997).

\bibitem{gardel-etal:2003}
M.~L. Gardel, M.~T. Valentine, J.~C. Crocker, A.~R. Bausch, and D.~A. Weitz,
  Phys.~Rev.~Lett. {\bf 91},  158302  (2003).

\bibitem{lutz-kollmann-bechinger:2004}
C. Lutz, M. Kollmann, and C. Bechinger, Phys.~Rev.~Lett. {\bf 93},  026001
  (2004).

\bibitem{pusey:91}
P.~N. Pusey,  in { Liquids, Freezing and Glass Transition}, edited by J.~P.
  Hansen, D. Levesque, and J. Zinn-Justin (Elsevier, North Holland, Amsterdam,
  1991), p.\ 763.

\bibitem{russel-saville-schowalter:91}
W.~B. Russel, D.~A. Saville, and W.~R. Schowalter, { Colloidal Dispersions}
  (Cambridge University Press, Cambridge, 1991).

\bibitem{frenkel:2000}
D. Frenkel,  in { Soft and Fragile Matter: Nonequilibrium Dynamics,
  Metastability and Flow}, edited by M.~E. Cates and M.~R. Evans (SUSSP {$\&$}
  IOP, Bristol, 2000), pp.\ 113--144.

\bibitem{hansen-mcdonald:86}
J.~P. Hansen and I.~R. McDonald, { Theory of simple liquids}, 2 ed.
  (Academic Press, London, 1986).

\bibitem{herzfeld:96}
J. Herzfeld, Acc. Chem. Res. {\bf 29},  31  (1996).

\bibitem{chaikin:2000}
P. Chaikin,  in { Soft and Fragile Matter: Nonequilibrium Dynamics,
  Metastability and Flow}, edited by M.~E. Cates and M.~R. Evans (SUSSP {$\&$}
  IOP, Bristol, 2000), pp.\ 315--348.

\bibitem{alder-wainwright:57}
B.~J. Alder and T.~E. Wainwright, J.~Chem.~Phys. {\bf 27},  1208  (1957).

\bibitem{hoover-ree:68}
W.~G. Hoover and F.~H. Ree, J.~Chem.~Phys. {\bf 49},  3609  (1968).

\bibitem{pusey-van_megen:86}
P.~N. Pusey and W. van Megen, Nature {\bf 320},  340  (1986).

\bibitem{zhu-etal:97}
J. Zhu, M. Li, R. Rogers, W. Meyer, R.~H. Ottewill, S. Space Shuttle~Crew,
  Russel, and C.~P. M., Nature {\bf 387},  883   (1997).

\bibitem{close-packing}
B. Cipra, Science {\bf 281},  1267  (1998).

\bibitem{bergfors:99}
T.~M. Bergfors, { Protein Crystallization} (International University Line,
  La Jolla, CA, 1999).

\bibitem{poon:97}
W.~C.~K. Poon, Phys.~Rev.~E {\bf 55},  3762  (1997).

\bibitem{einstein:05p}
A. Einstein, Ann. Phys. (Leipzig) {\bf 17},  132  (1905).

\bibitem{john:91}
S. John, Physics Today {\bf 44},  32  (1991).

\bibitem{goetze:91}
W. G{\"o}tze,  in { Liquids, Freezing and Glass Transition}, edited by J.~P.
  Hansen, D. Levesque, and J. Zinn-Justin (Elsevier, North Holland, Amsterdam,
  1991), p.\ 287.

\bibitem{Fuchs94}
M. Fuchs, J.~Non-Cryst.~Solids {\bf 172--174},  241  (1994).

\bibitem{Goetze92}
W. G\"otze and L. Sj\"ogren, Rep.~Prog.~Phys. {\bf 55},  241  (1992).

\bibitem{Goetze99}
W. G\"otze, J.~Phys.: Condens.~Matter {\bf 11},  A1  (1999).

\bibitem{van_megen-underwood:93}
W. van Megen and S.~M. Underwood, Phys.~Rev.~Lett. {\bf 70},  2766  (1993).

\bibitem{kob:99}
W. Kob, J.~Phys.:~Condens.~Matter {\bf 11},  R85  (1999).

\bibitem{cates:2003}
M.~E. Cates, Ann. Henri Poincar{\'e} {\bf 4},  S647  (2003).

\bibitem{bouchaud-etal:96}
J.-P. Bouchaud, L. Culiandolo, J. Kurchan, and M. Mezard, Physica A {\bf 226},
  243  (1996).

\bibitem{Kawasaki98}
K. Kawasaki and K. Fuchizaki, J.~Non-Cryst.~Solids {\bf 235--237},  57  (1998).

\bibitem{pham-etal:2002}
K.~N. Pham, A.~M. Puertas, J. Bergenholtz, S.~U. Egelhaaf, A. Moussaid, P.~N.
  Pusey, A.~B. Schofield, M.~E. Cates, M. Fuchs, and W.~C.~K. Poon, Science
  {\bf 296},  104  (2002).

\bibitem{sciortino:2002}
F. Sciortino, Nature Mat. {\bf 1},  145  (2002).

\bibitem{brochard-lennon:75}
F. Brochard and J.~F. Lennon, J.~Phys.~(Paris) {\bf 36},  1035  (1975).

\bibitem{helfrich-harbich:85}
W. Helfrich and W. Harbich, Chem.~Scr. {\bf 25},  32  (1985).

\bibitem{mackintosh-kaes-janmey:95}
F. MacKintosh, J. K{\"a}s, and P. Janmey, Phys.~Rev.~Lett. {\bf 75},  4425
  (1995).

\bibitem{hinner-etal:98}
B. Hinner, M. Tempel, E. Sackmann, K. Kroy, and E. Frey, Phys.~Rev.~Lett. {\bf
  81},  2614  (1998).

\bibitem{halpin-healy_zhang:95}
T. Halpin-Healy and Y.-C. Zhang, Phys. Rep. {\bf 254},  215  (1995).

\bibitem{kardar:98}
M. Kardar, Phys. Rep. {\bf 301},  85  (1998).

\bibitem{fisher:98}
D. Fisher, Phys.~Rep. {\bf 301},  113  (1998).

\bibitem{panja:2004}
D. Panja, Phys.~Rep. {\bf 393},  87  (2004).

\bibitem{degennes:79}
P.~G. de~Gennes, { Scaling Concepts in Polymer Physics} (Cornell University
  Press, Ithaca and London, 1979).

\bibitem{safran:2003}
S.~A. Safran, { Statistical Thermodynamics of Surfaces, Interfaces, and
  Membranes} (Westview Press, Boulder, 2003).

\bibitem{seifert:97}
U. Seifert, Adv. Phys. {\bf 46},  13  (1997).

\bibitem{lipowsky:91}
R. Lipowsky, Nature {\bf 349},  475  (1991).

\bibitem{lipowsky_sackmann:95a}
{ Structure and Dynamics of Membranes: From Cells to Vesicles. Handbook of
  Biological Physics, Vol. 1.}, edited by R. Lipowsky and E. Sackmann (Elsevier
  Science, Amsterdam, 1995).

\bibitem{nelson:book}
D. Nelson, { Defects {$\&$} Geometry in Condensed Matter Physics}, 1 ed.
  (Cambridge University Press, Cambridge, 2002).

\bibitem{blatter:review}
G. Blatter, M. Feigelman, V. Geshkenbein, A. Larkin, and V. Vinokur,
  Rev.~Mod.~Phys. {\bf 66},  1125  (1994).

\bibitem{frey_nelson:91}
E. Frey and D. Nelson, J. Phys. I (France) {\bf 1},  1715  (1991).

\bibitem{gittes-etal:93}
F. Gittes, B. Mickey, J. Nettleton, and J. Howard, Journal of Cell Biology {\bf
  120},  923  (1993).

\bibitem{caspi-etal:98}
A. Caspi, M. Elbaum, R. Granek, A. Lachish, and D. Zbaida, Phys.~Rev.~Lett.
  {\bf 80},  1106  (1998).

\bibitem{legoff-etal:2002}
L. Le~Goff, O. Hallatschek, E. Frey, and F. Amblard, Phys.~Rev.~Lett. {\bf 89},
   258101  (2002).

\bibitem{hohenadl-etal:99}
M. Hohenadl, T. Storz, H. Kirpal, K. Kroy, and R. Merkel, Biophys. J. {\bf 77},
   2199  (1999).

\bibitem{hallatschek-frey-kroy:2004}
O. Hallatschek, E. Frey, and K. Kroy, to be published (unpublished).

\bibitem{wiese:2000}
K.~J. Wiese,  in { Polymerized membranes, a review}, Vol.~19 of { Phase
  Transitions and Critical Phenomena}, edited by C. Domb (Academic Press,
  Oxford, 2000), Chap.~2.

\bibitem{mcleish:2002}
T.~C.~B. McLeish, Adv. Phys. {\bf 51},  1379   (2002).

\bibitem{kaes-strey-sackmann:94}
J. K{\"a}s, H. Strey, and E. Sackmann, Nature {\bf 368},  226  (1994).

\bibitem{smith-perkins-chu:95}
D.~E. Smith, T.~T. Perkins, and S. Chu, Phys.~Rev.~Lett. {\bf 75},  4146
  (1995).

\bibitem{chong-fuchs:2002}
S.~H. Chong and M. Fuchs, Phys.~Rev.~Lett. {\bf 88},  185702  (2002).

\bibitem{everaers-etal:2004}
R. Everaers, S.~K. Sukumaran, G.~S. Grest, C. Svaneborg, A. Sivasubramanian,
  and K. Kremer, Science {\bf 303},  823  (2004).

\bibitem{wittmer-etal:2004}
J.~P. Wittmer, H. Meyer, J. Baschnagel, A. Johner, S. Obukhov, L. Mattioni, M.
  Müller, and A.~N. Semenov, Phys.~Rev.~Lett. {\bf 93},  147801  (2004).

\bibitem{ramaswamy-toner-prost:2000}
S. Ramaswamy, J. Toner, and J. Prost, Phys.~Rev.~Lett. {\bf 84},  3494  (2000).

\bibitem{manneville_etal:99}
D.~L. J.-B.~Manneville, P.~Basserau and J. Prost, Phys.~Rev.~Lett. {\bf 82},
  4356  (1999).

\bibitem{nedelec-etal:97}
F.~J. Nedelec, T. Surrey, A.~C. Maggs, and S. Leibler, Nature {\bf 389},  305
  (1997).

\bibitem{humphrey_etal:02}
D. Humphrey, C. Duggan, D. Saha, D. Smith, and J. K{\"a}s, Nature {\bf 416},
  413  (2002).

\bibitem{uhde_etal:04}
J. Uhde, M. Keller, E. Sackmann, A. Parmeggiani, and E. Frey, Phys.~Rev.~Lett.
  {\bf (in press)},    (2004).

\bibitem{kruse-etal:2004}
K. Kruse, J.~F. Joanny, F. Jülicher, J. Prost, and K. Sekimoto,
  Phys.~Rev.~Lett. {\bf 92},  078101  (2004).

\bibitem{}
P. H{\"a}nggi, P. Talkner, and m.~Borkovec, Rev.~Mod.~Phys. {\bf 62},  251
  (1990).

\bibitem{hatwalne-etal:2004}
Y. Hatwalne, S. Ramaswamy, M. Rao, and R.~A. Simha, Phys.~Rev.~Lett. {\bf 92},
  118101  (2004).

\bibitem{helmholtz:lebendige_kraft}
H. v.~Helmholtz, { in Ostwalds Klassiker der Exakten Wissenschaften} (Harri
  Deutsch, Thun, 1996), Vol.~1.

\bibitem{huxley:review}
A. Huxley, Prog. Biophys. {\bf 7},  255  (1957).

\bibitem{hill:review}
T. Hill, Prog. Biophys. Mol. Biol. {\bf 28},  267  (1974).

\bibitem{mehta99a}
A.~D. Mehta, M. Rief, J.~A. Spudich, D.~A. Smith, and R.~M. Simmons, Science
  {\bf 283},  1689  (1999).

\bibitem{howard:book}
J. Howard, { Mechanics of Motor Proteins and the Cytoskeleton} (Sinauer
  Associates, Sunderland, MA, 2001).

\bibitem{haenggi_bartussek:96}
P. H{\"a}nggi and R. Bartussek, Lect. Notes Phys. {\bf 476},  294  (1996).

\bibitem{astumian:review}
R. Astumian, Science {\bf 276},  917  (1997); R. Astumian and
P. H\"anggi, Physics Today {\bf 55}, 33 (2002).

\bibitem{juelicher_etal:review}
F. J{\"u}licher, A. Ajdari, and J. Prost, Rev.~Mod.~Phys. {\bf 69},  1269
  (1997).

\bibitem{reimann:review}
P. Reimann, Phys. Rep. {\bf 361},  57  (2002).

\bibitem{parmeggiani_schmidt:review}
A. Parmeggiani and C. Schmidt,  in { Functions and Regulation of Cellular
  Systems: Experiments and Models}, edited by A. Deutsch, J. Howard, M. Falcke,
  and W. Zimmermann (Birkh{\"a}user, Basel, 2004), p.\ 151.
  
\bibitem{haenggi_atal:05} P. H{\"a}nggi, F. Marchesoni, and F. Nori,
  Annalen der Physik {\bf 14}, xxxx (2005); cond-mat/0410033.

\bibitem{bug_berne:87}
A. Bug and B. Berne, Phys.~Rev.~Lett. {\bf 59},  948  (1987).

\bibitem{ajdari92}
A. Ajdari and J. Prost, C.R. Acad.~Sci.~Paris~II {\bf 315},  1635  (1992).

\bibitem{astumian94}
R.~D. Astumian and M. Bier, Phys.~Rev.~Lett. {\bf 72},  1766  (1994).

\bibitem{smoluchowski:1912}
M. von Smoluchowski, Phys. Zeitschr. {\bf 13},  1069  (1912).

\bibitem{rousselet_etal:94}
J. Rousselet, L. Salome, A. Ajdari, and J. Prost, Nature {\bf 370},  446
  (1994).

\bibitem{faucheux_etal:95}
L. Faucheux, L. Bourdieu, P. Kaplan, and A. Libchaber, Phys.~Rev.~Lett. {\bf
  74},  1504  (1995).

\bibitem{sarikay_etal:03}
M. Sarikay, C. Tamaeler, A. Jen, K. Schulten, and F. Baneyx, Nature Materials
  {\bf 2},  577  (2003).

\bibitem{svoboda94b}
K. Svoboda, P.~P. Mitra, and S.~M. Block, Proc.\ Natl.\ Acad.\ Sci.\ (USA) {\bf
  91},  11782  (1994).

\bibitem{macdonald_gibbs_pipkin:68}
C. MacDonald, J. Gibbs, and A. Pipkin, Biopolymers {\bf 6},  1  (1968).

\bibitem{spohn:book}
H. Spohn, { Large Scale Dynamics of Interacting Particles} (Springer Verlag,
  New York, 1991).

\bibitem{derrida_evans:review}
B. Derrida and M. Evans,  in { Nonequilibrium Statistical Mechanics in One
  Dimension}, edited by V. Privman (Cambridge University Press, Cambridge,
  1997), Chap.~14, pp.\ 277--304.

\bibitem{mukamel:review}
D. Mukamel,  in { Soft and Fragile Matter}, edited by M. Cates and M. Evans
  (Institute of Physics Publishing, Bristol, 2000), pp.\ 237--258.

\bibitem{schuetz:review}
G. Sch{\"u}tz,  in { Phase Transitions and Critical Phenomena}, edited by C.
  Domb and J. Lebowitz (Academic Press, San Diego, 2001), Vol.~19, pp.\ 3--251.

\bibitem{alberts}
B. Alberts, D. Bray, J. Lewis, M. Raff, K. Roberts, and J.~D. Watson, {
  Molecular Biology of the cell}, $3^{\rm rd}$ ed. (Garland Publ., New York,
  1994).

\bibitem{beijeren_kutner_spohn:85}
H. van Beijeren, R. Kutner, and H. Spohn, Phys.~Rev.~Lett. {\bf 54},  2026
  (1985).

\bibitem{meakin_etal:86}
P. Meakin, P. Ramanlal, L. Sander, and R. Ball, Phys.~Rev.~A {\bf 34},  5091
  (1986).

\bibitem{kardar_parisi_zhang:86}
M. Kardar, G. Parisi, and Y.-C. Zhang, Phys.~Rev.~Lett. {\bf 56},  889  (1986).

\bibitem{forster-nelson-stephen:77}
D. Forster, D.~R. Nelson, and M.~J. Stephen, Phys.~Rev.~A {\bf 16},  732
  (1977).

\bibitem{krug:91}
J. Krug, Phys.~Rev.~Lett. {\bf 67},  1882  (1991).

\bibitem{landau_lifshitz_stat_mech:book}
L. Landau and E. Lifshitz, { Statistical Physics I} (Pergamon Press, New
  York, 1980).

\bibitem{derrida_lebowitz_speer:01}
B. Derrida, J. Lebowitz, and E. Speer, Phys.~Rev.~Lett. {\bf 87},  150601
  (2001).

\bibitem{derrida_lebowitz_speer:02}
B. Derrida, J. Lebowitz, and E. Speer, Phys.~Rev.~Lett. {\bf 89},  030601
  (2002).

\bibitem{shaw_zia_lee:03}
L. Shaw, R. Zia, and K. Lee, Phys.~Rev.~E {\bf 68},  021910  (2003).

\bibitem{chou:03}
T. Chou,  {\bf 85},  755  (2003).

\bibitem{parmeggiani_franosch_frey:03}
A. Parmeggiani, T. Franosch, and E. Frey, Phys.~Rev.~Lett. {\bf 90},  068810
  (2003).

\bibitem{klumpp_lipowsky:03}
S. Klumpp and R. Lipowsky, J. Stat. Phys. {\bf 113},  233  (2003).

\end{thebibliography}
\end{document}